\documentclass[useAMS,usenatbib]{mn2e}
\usepackage{epsfig,color}

\usepackage{natbib}
\bibpunct{(}{)}{;}{a}{}{,}
\def\inter{\rm{int}}
\def\cap{\rm{cap}}

\def\spose#1{\hbox to 0pt{#1\hss}}
\def\ltsimm{\mathrel{\spose{\lower 3pt\hbox{$\sim$}}
        \raise 2.0pt\hbox{$<$}}}
\def\gtsimm{\mathrel{\spose{\lower 3pt\hbox{$\sim$}}
        \raise 2.0pt\hbox{$>$}}}

\def\km{{\rm\thinspace km}}
\def\cm{{\rm\thinspace cm}}

\def\s{{\rm\thinspace s}}
\def\yr{{\rm\thinspace yr}}
\def\g{{\rm\thinspace g}}
\def\kmps{\hbox{${\rm\km\s^{-1}\,}$}}

\def\erg{{\rm\thinspace erg}}
\def\Hz{{\rm\thinspace Hz}}

\def\ster{{\rm\thinspace ster}}
\def\ergps{\hbox{${\rm\erg\s^{-1}\,}$}}
\def\Rsol{\hbox{${\rm\thinspace R_{\odot}}$}}
\def\Msol{\hbox{${\rm\thinspace M_{\odot}}$}}

\def\Msolpyr{\hbox{${\rm\Msol\yr^{-1}\,}$}}
\def\pcm{\hbox{${\rm\cm^{-1}\,}$}}
\def\pcm2{\hbox{${\rm\cm^{-2}\,}$}}
\def\pcm3{\hbox{${\rm\cm^{-3}\,}$}}
\def\ergpscm3Hz{\hbox{${\rm\ergps\cm^{-3}\Hz^{-1}\,}$}}
\def\ergpscm3Hzster{\hbox{${\rm\ergps\cm^{-3}\Hz^{-1}\ster^{-1}\,}$}}
\def\gpcm3{\hbox{${\rm\g\cm^{-3}\,}$}}
\def\ergpcm2{\hbox{${\rm\erg\cm^{-2}\,}$}}
\def\ergpcm3{\hbox{${\rm\erg\cm^{-3}\,}$}}
\def\phpscm2{\hbox{${\rm photons\s^{-1}\cm^{-2}\,}$}}

\def\etacar{$\eta\thinspace\rm{Car}\thinspace$}

\title[A 3D dynamical model of the colliding
winds in binary systems]{A 3D dynamical model of the colliding
winds in binary systems}

\author[E.~R.~Parkin \& J.~M.~Pittard] {E. R. Parkin\thanks{E-mail:
phy1erp@leeds.ac.uk} \& J.~M.~Pittard\\ School of Physics and
Astronomy, The University of Leeds, Leeds LS2 9JT, UK}

\begin{document}

\date{Accepted 2008 May 21. Received 2008 May 20; in original form
2008 April 22}

\pagerange{\pageref{firstpage}--\pageref{lastpage}} \pubyear{2008}

\maketitle

\label{firstpage}

\begin{abstract}
We present a 3D dynamical model of the orbital induced curvature of
the wind-wind collision region in binary star systems. Momentum
balance equations are used to determine the position and shape of the
contact discontinuity between the stars, while further downstream the
gas is assumed to behave ballistically. An archimedean spiral
structure is formed by the motion of the stars, with clear resemblance
to high resolution images of the so-called ``pinwheel nebulae''. A key
advantage of this approach over grid or smoothed particle hydrodynamic
models is its significantly reduced computational cost, while it also
allows the study of the structure obtained in an eccentric orbit. The
model is relevant to symbiotic systems and $\gamma$-ray binaries, as
well as systems with O-type and Wolf-Rayet stars.

As an example application, we simulate the X-ray emission from
hypothetical O+O and WR+O star binaries, and describe a method of ray
tracing through the 3D spiral structure to account for absorption by
the circumstellar material in the system. Such calculations may be
easily adapted to study observations at wavelengths ranging from the
radio to $\gamma$-ray.
\end{abstract}

\begin{keywords}
hydrodynamics - methods:numerical - stars:early-type - X-rays:stars -
stars:binaries - stars:winds
\end{keywords}

\section{Introduction}
\label{sec:intro}
Colliding winds occur in various types of stellar binaries, including
those with massive OB and Wolf-Rayet (WR) stars, lower mass eruptive
symbiotic systems containing a white dwarf and red giant star which
undergo a ``slow nova'' outburst, and binary systems which contain one
or two pulsars blowing a pulsar wind(s).

High-spatial-resolution observations are revealing many interesting
features in such systems. In massive O+O and WR+O binaries, radio
interferometry has spatially-resolved emission from non-thermal
electrons at the apex of the wind-wind collision
\citep[e.g.,][]{Williams:1997,Dougherty:2000,Dougherty:2005,Contreras:2004}.
Beautiful ``pinwheel'' structures which trace dust emission can also
be observed \citep[e.g.,][]{Tuthill:1999,Tuthill:2006,
Tuthill:2008,Monnier:1999,Marchenko:2002}. The shape of these
structures can be described by archimedean spirals which are believed
to follow the wind-wind collision region in systems where the winds
are of very unequal momentum.

Colliding winds also play a key role in eruptive symbiotic systems,
where a hot, fast, diffuse wind from a white dwarf companion interacts
with a slow massive wind from a Mira type primary star\footnote{Such
systems are to be distinguished from the interacting wind phenomenon
which occurs in AGB binaries where density structures in the AGB wind
are created either due to the reflex-action of the evolved star around
the centre-of-mass of the system \citep{Mastrodemos:1999,
Mauron:2006,He:2007}, or due to gravitational focusing
\citep{Gawryszczak:2002}. Here we focus exclusively on systems
involving the interaction of winds from separate stars.}. The class of
eruptive symbiotics can be divided into two further subtypes:
classical symbiotics, in which the bolometric luminosity remains
constant and outbursts typically last about 100 days (Z~And is an
example), and the more powerful eruptions known as symbiotic novae,
where the bolometric luminosity increases by a factor of order 10-100
on a timescale of about a year, and the system stays in an active
state for $\gtsimm 10\;$yrs (well-known examples are V1016~Cyg,
HM~Sge, and AG~Pegasi).

Colliding winds may also play a key role in the newly discovered class
of systems called $\gamma$-ray binaries \citep{Aharonian:2005a,
Aharonian:2005b, Albert:2006}. The nature of these systems is still
controversial, though in the case of PSR B1259-63, it is clear that a
relativistic wind from a pulsar collides with the stellar wind from a
Be star. The orbit is highly eccentric ($e = 0.87$), and has a period
of 3.4 yr \citep{Johnston:2005}. The TeV $\gamma$-ray emission arises
from the Inverse Compton cooling of ultra-relativistic electrons
accelerated at the pulsar wind termination shock
\citep[e.g.][]{Khangulyan:2007}. In contrast, the nature of the
sources LS5039 and LS I +61 303 is less clear, since the type of
compact object has not been established beyond doubt
\citep{Romero:2007,Dubus:2008,Khangulyan:2008}.

While there has been much progress in modelling the dynamical
structure of the colliding winds in early-type binary systems, the
majority of work has been limited to 2D \citep*[e.g.,
][]{Stevens:1992,
Gayley:1997,Pittard:1997,Pittard:1998,Pittard:2007,Zhekov:2007}.
3-dimensional hydrodynamical calculations have been performed by
\citet{Pittard:1999}, \citet{Walder:2002}, and \citet{Lemaster:2007},
while a ballistic model was presented by \citet{Harries:2004}. An SPH
model has recently been computed by \citet{Okazaki:2008}. Dynamical
models for symbiotic novae have been presented by \citet{Girard:1987}
and \citet{Kenny:2005,Kenny:2007}, while 3D hydrodynamical models have
been presented by \citet{Walder:2000}. Models of the wind-wind
collision in classical symbiotics have been presented by
\citet{Mitsumoto:2005} and \citet{Bisikalo:2006}. Relativistic
hydrodynamics \citep{Bogovalov:2008} and SPH \citep{Romero:2007}
models have been used to investigate the wind-wind collision in pulsar
wind binary systems.

Although dramatic improvements in computational power and techniques
in recent years have spurred the development of 3D models of colliding
winds, such work remains computationally expensive, and it is still
difficult to perform simulations of CWB's even on high performance
parallel machines when the orbital eccentricity is high. We therefore
present a new method which captures the flow dynamics while requiring 
less computational resources.

At its heart, our approach adopts the equations for the ram pressure
balance between the two winds as detailed by \citet{Canto:1996}. In
this work it is assumed that both winds are highly radiative, rapidly
cool, and fully mix. While these assumptions are only relevant in
close binaries, it provides a convenient starting point and
the position of the contact discontinuity is unlikely to
drastically change even if the wind-wind collision is essentially
adiabatic. Then, at some distance downstream of the apex of the
wind-wind collision region (WCR) the flow is assumed to reach a
terminal speed and to thereafter flow ballistically (i.e. no net force
acting upon it). This ballistic treatment has similarities to many
previous works
\citep[e.g.,][]{Girard:1987,Harries:2004,Kenny:2007}. The derivation in
\citet{Canto:1996} has also been widely used to model observable
properties \citep[e.g.,][]{Foellmi:2008,Henley:2008}.

This paper is organised as follows. In \S~\ref{sec:shkconemodel} we
explain the steps necessary to construct our dynamical model of the
wind-wind collision. \S~\ref{subsec:emission} shows how it can be used
to simulate the X-ray emission and circumstellar absorption arising
from the WCR in early-type binary systems, though this is but one
example of the potential use of such a model.  In
\S~\ref{sec:conclusions} we summarize and conclude our findings, and
outline possible future directions.

\section{The dynamical model}
\label{sec:shkconemodel}

\subsection{Overview}
\label{subsec:shkconeoverview}

In the model the orbit is calculated in the frame of one of the stars
(herafter referred to as the primary star). The winds are assumed to reach
their terminal speeds before they collide.
The contact discontinuity (CD) is split into two sections, to account 
for the effect of orbital motion:

i) A region close to the apex of the WCR where the flow from the
stagnation point is accelerating along the CD (hereafter called the
``shock cap''). The shock cap is terminated where the flow is assumed
to become ballistic (the ``ballistic point''), the exact point being
calibrated against hydrodynamical models (see
\S~\ref{subsec:bcd}). While the properties of the shock cap are
assumed to be axisymmetric, orbital motion introduces an aberration
angle which means that the symmetry axis and the line of centres of
the stars are not colinear (see \S~\ref{subsec:shockcap}).

ii) A region beyond the ballistic point where the flow along the
contact discontinuity is unaffected by the primary and secondary
stars' gravity, ram pressure from the winds, or thermal pressure in
the WCR. If the stellar winds have differing speeds, the flow in this
region is assumed to move with the speed of the slower wind, since
this is the wind which responds least to the orbital motion of the
stars, and dominates the absorption in the system (in the models
presented in this paper, both winds have the same speed of $2000
\thinspace \rm{kms}^{-1}$ - see Table~\ref{tab:models}). This
region is termed the ``ballistic CD''. 

\begin{figure}
\begin{center}
\psfig{figure=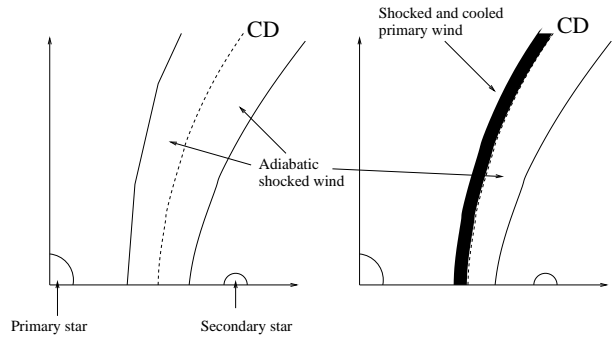,width=8.0cm}
\caption[]{Schematic diagram showing the location of the shocks
bounding the CD for varying values of the cooling parameter. Left:
both winds have $\chi \gtsimm 1$ and are therefore largely
adiabatic. Right: The primary wind has $\chi \ll 1$ and is strongly
radiative, whereas the secondary's wind is adiabatic. The scenario on
the left represents the O+O and WR+O binaries investigated in this
paper, whereas the scenario on the right represents what is thought to
occur in \etacar.}
\label{fig:wcr}
\end{center}
\end{figure}

By separating the CD into these two sections we can model the effect
of the winding of the CD around the stars and the subsequent
absorption by the un-shocked winds. We do not attempt to model the
shocks which bound either side of the CD in this work, as in many
circumstances the shocked gas efficiently cools and is compressed by
the ram pressure of the pre-shock wind into a thin dense sheet
coincident with the CD.  For instance, in symbiotic novae, the hot
wind is likely to be strongly radiative \citep[see Fig.~4
in][]{Kenny:2005}, as of course is the cool wind, and our model
therefore gives the position of the shocked gas and the dense spiral
shells which subsequently form. Strong radiative cooling is also a
feature of the WCR in many massive binaries.  In \etacar, for example,
the primary LBV wind is so dense (and slow) that it is strongly
radiative around the entire orbit \citep{Pittard:1998b}.  The
importance of cooling in the WCR can be quantified using the cooling
parameter \citep{Stevens:1992},

\begin{equation}
   \chi = \frac{t_{\rm{cool}}}{t_{\rm{esc}}} = \frac{v^4 _8
   d_{12}}{\dot{M} _{-7}},
\end{equation}

\noindent where $v_8$ is the wind velocity in units of $1000
\thinspace\rm{km\thinspace s}^{-1}$, $d_{12}$ is the separation of the
stars in units of $10^{12}\rm{cm}$, $\dot{M}_{7}$ is the mass-loss
rate of the star in units of $10^{-7} \Msolpyr$, $t_{\rm{cool}}$ is
the cooling time, and $t_{\rm{esc}}$ ($=d / c_{\rm s}, \thinspace
c_{\rm s}$ is the postshock sound speed) is the characteristic time
for hot gas near the apex of the WCR to flow downstream. In practice,
hydrodynamical simulations show that the wind collision region (WCR)
is adiabatic for $\chi \gtsimm 3$, whereas for $\chi \ltsimm 3$ it
cools rapidly (Fig.~\ref{fig:wcr}). 

\begin{figure}
\begin{center}
\psfig{figure=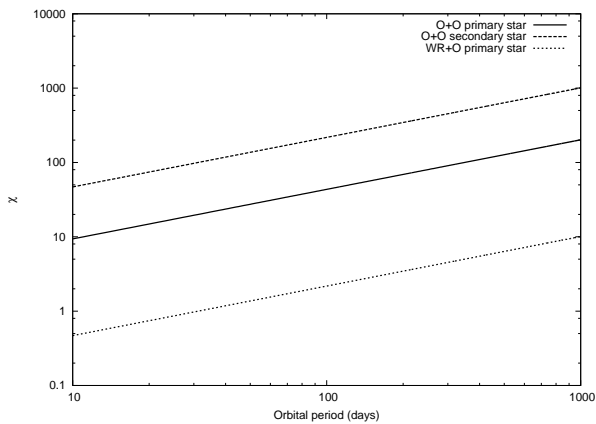,width=8.0cm}
\caption[]{Ratio of the cooling to characteristic flow timescale
for the hot shocked gas in a colliding winds binary as a function of
orbital period. Cooling becomes important once $\chi \ltsimm 3$.}
\label{fig:chi}
\end{center}
\end{figure}

Fig.~\ref{fig:chi} shows the value of $\chi$ as a function of orbital
period for each shocked wind in a hypothetical O+O star binary with a
circular orbit, wind speeds of $2000\;\kmps$, and mass-loss rates of
$10^{-6}\;\Msolpyr$ and $2 \times 10^{-7}\;\Msolpyr$ for the primary
and secondary star of masses 50 and 30 \Msol\space
respectively. Clearly both of the shocked winds are largely adiabatic,
even down to an orbital period of $10\;$days (in shorter period
systems the stars are close enough together that
acceleration/deceleration of the winds needs to be considered).
However, in a hypothetical WR+O system where the mass-loss rates of
the primary and secondary stars are now $2 \times 10^{-5}\;\Msolpyr$
and $10^{-6}\Msolpyr$, and both stars have masses of $50\Msol$ and
wind speeds of $2000\thinspace \rm{kms}^{-1}$, cooling is important
for orbital periods $\ltsimm 1\;$yr. If the WR star is a WC subtype,
cooling is important for periods up to several years, since cooling is
more efficient with such abundances \citep[see,
e.g.,][]{Stevens:1992}.  Thus, Fig.~\ref{fig:chi} shows that the
denser winds from WR stars are likely to produce radiative shocks in
many instances, though the O+O systems will usually be adiabatic
unless the orbital period, $P \ltsimm 10\;$d, or the winds are slower
and/or denser than assumed above.

The postshock winds of both the primary and secondary stars in the
simulations discussed in \S~\ref{subsec:emission} are largely
adiabatic.  In such cases, the temperature of the hot gas in the WCR
as a function of distance downstream from the stagnation point at the
apex of the WCR has been determined by \citet{Kenny:2005}. With this
information it is possible to derive the width of the post-shock
layer, and hence the position of the shocks, as a function of
downstream distance. However, this is beyond the scope of the present
work.
 
In the following sections we detail the modelling of the shock cap and
ballistic CD.

\subsection{The shock cap}
\label{subsec:shockcap}

\begin{figure}
\begin{center}
\psfig{figure=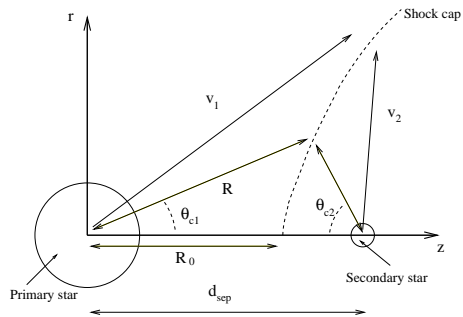,width=6.0cm}
\caption[]{Schematic diagram describing the wind-wind interaction
between the two stars represented by circles.}
\label{fig:cantofig1}
\end{center}
\end{figure}

The shape of the shock cap is determined from momentum balance
requirements. The surface density and velocity of the flow along the
shock cap are obtained from Eqs.~29 and 30 of \cite{Canto:1996} (the
latter scaled to the speed of the slower wind). Assuming the winds are
already at their terminal velocity when they reach the shocks, the
locus of the CD is $R(\theta_{\rm c1})$, where $R$ is the distance
from the centre of the primary star and $\theta_{\rm c1}$ is the angle
between the vector to the primary star and the line-of-centres (see
Fig.~\ref{fig:cantofig1}). The ratio of the wind momenta is given by

\begin{equation}
   \eta \equiv \frac{\dot{M}_{2}v_{\infty2}}{\dot{M}_{1}v_{\infty1}},
\label{eqn:eta}
\end{equation}

\noindent where $\dot{M_{1}}$, $v_{\infty1}$, $\dot{M_{2}}$, and
$v_{\infty2}$ are the mass-loss rates and terminal velocities of the
primary and secondary stars respectively.

The shock cap is symmetrical about the line of centres before the
effects of orbital motion are introduced. The 2D ($r,z$) coordinates of
points on the shock cap in units of the stellar separation, $d_{\rm sep}$,
are

\begin{equation}
   z =  \frac{\tan\theta_{c_1}}{\tan\theta_{\rm c2}+\tan\theta_{\rm c1}},\\
\label{eqn:zandr1}
\end{equation}
\begin{equation}
   r  =  z\tan\theta_{\rm c2}. 
\label{eqn:zandr2}
\end{equation}

\noindent To determine the coordinates in 3D, the 2D arms of the WCR can be
rotated azimuthally. The $x, y$, and $z$
vectors ($x_{\rm{cap}}$, $y_{\rm{cap}}$, and
$z_{\rm{cap}}$ respectively) from the center of the
primary star to coordinates on the shock cap are then

\begin{eqnarray*}
   x_{\rm{cap}} =  & d_{\rm{sep}}(z\cos\omega - r\sin\omega\cos\zeta), \\
   y_{\rm{cap}} =  & d_{\rm{sep}}(z\sin\omega + r\cos\omega\cos\zeta), \\
   z_{\rm{cap}} =  & d_{\rm{sep}}(r\sin\zeta),
\end{eqnarray*} 

\noindent where $\omega$ is the true anomaly of the orbit and $\zeta$
is the azimuthal angle subtended between a coordinate on the surface
of the shock cap, the line of centres, and the $xy$ (orbital) plane.

The number of coordinate points on the shock cap is determined by the
values of $\theta_{\rm c1\infty}$, $\delta\theta_{c1}$ and
$\delta\zeta$. With $\delta\theta_{\rm c1} = 1^{\circ}$ and
$\delta\zeta = 18^{\circ}$ (i.e. 20 azimuthal points per 2D $rz$
value), the shock cap consists of $\sim 10^{3}$ separate coordinate
points.  Eq.~29 of \cite{Canto:1996} is used to determine the
tangential velocity along the CD, and thus the position of the
ballistic point in 2D axisymmetry.

\begin{figure}
\begin{center}
\psfig{figure=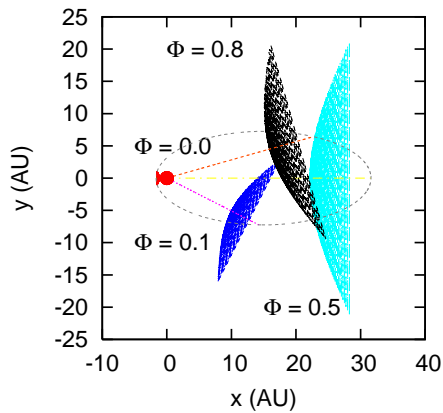,width=8.0cm}
\caption[]{Plan view of the shock cap at various orbital phases in a
system with an orbital eccentricity, $e=0.9$. The calculations are
performed in the frame of the primary star (located at the origin and
not to scale) with the secondary star orbiting in an anti-clockwise
direction. As the separation of the stars increases the linear size of
the shock cap also increases in direct proportion. This is most
noticeable when comparing the shock cap at $\phi=0.0$ (periastron,
red) and $\phi=0.5$ (apastron, turquoise). The aberration (skew) of
the shock cap due to orbital motion is not considered in this plot.}
\label{fig:shkcones}
\end{center}
\end{figure}

Since the size of the wind-wind collision scales with the orbital
separation, dramatic variations occur in systems with highly eccentric
orbits, as shown in Fig.~\ref{fig:shkcones} where $e=0.9$; the high
eccentricity means that the shock cap at periastron has a linear scale
which is 20 times smaller than that at apastron. Such high
eccentricities occur in two of the most well-known colliding winds
systems, \etacar and WR\thinspace140, and also in PSR B1259-63, one of
the $\gamma$-ray binaries.

Another effect resulting from orbital motion is the aberration (skew)
of the apex of the WCR due to the net velocity vector of the orbit
(i.e. the motion of the secondary star relative to the primary star).
The skew angle, $\mu$, which is the angle between the symmetry axis of
the shock cap and the line of centres of the stars is approximated by
\begin{equation}
\tan\mu = \frac{v_{\rm{orb}}}{v_{\infty}},
\label{eqn:coriolis}
\end{equation}
\noindent where the speed of the slower wind is used. 
In the frame of the primary star, 
\begin{equation}
v_{\rm{orb}} = \left[G(M_{1} + M_{2})\left(\frac{2}{d_{\rm{sep}}}-\frac{1}{a}\right)\right]^{1/2},
\label{eqn:coriolis2}
\end{equation}
\noindent for stars of mass $M_{1}$ and $M_{2}$ and an orbital
semi-major axis, $a$. 

The aberration is significant in symbiotic novae because of the low
wind speed of the cool star (for instance, a symbiotic system with $e
= 0.0$, $M_{1}+M_{2}=2.5\;\Msol$, and $d_{\rm sep}=10\;$au has $v_{\rm
orb} = 15\;\kmps$, which is comparable to the speed of the cool
wind). In contrast, the aberration is small in early-type binaries
(for instance, an O+O binary with $e=0.0$, $M_{1}+M_{2}=80\;\Msol$,
and $d_{\rm sep}=4.3\;$au has $v_{\rm orb} = 130\;\kmps$, which is
typically much smaller than the wind speeds), unless the orbit has
high eccentricity. In such cases the magnitude of the skew varies
throughout the orbit.

\begin{figure}
\begin{center}
\psfig{figure=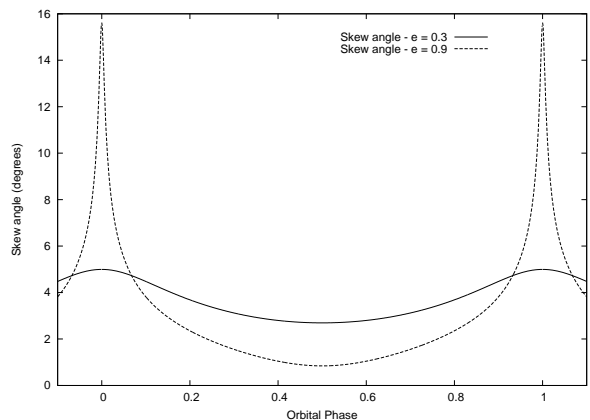,width=8.0cm}
\caption[]{Variation of the skew angle of the shock cap due to orbital
motion for eccentricities of $e=0.3$ and $e=0.9$, with $P = 1\;$yr and
$M_{1}+M_{2}=80\;\Msol$. The angle peaks at phase $\phi$ = 0.0, when
the stars are at closest approach and their relative orbital velocity
is a maximum.}
\label{fig:skew_angle}
\end{center}
\end{figure}

\begin{figure}
\begin{center}
\psfig{figure=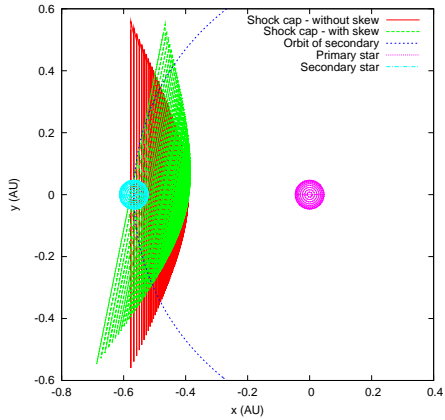,width=8.0cm}
\caption[]{Plan view of the shock cap at $\phi$ = 0.0 for $e = 0.9$,
$P = 1\;$yr and $M_{1}+M_{2}=80\;\Msol$, with (green) and without
(red) taking account of orbit induced skew. The secondary star orbits
in the anti-clockwise direction. Note that the shocks either side of
the CD are not displayed.}
\label{fig:shkcap_skew}
\end{center}
\end{figure}

Fig.~\ref{fig:skew_angle} shows how $\mu$ varies throughout the orbit
for the Model A O+O star binary with parameters as in
Table~\ref{tab:models} and with $e=0.3$ and 0.9. A peak value is
reached at periastron passage ($\phi = 0.0$) when the relative orbital
speed of the stars reaches it's highest value, and the lowest value of
$\mu$ occurs when the stars are at apastron ($\phi = 0.5$) and the
relative orbital velocity is a minimum. The variation of $\mu$ between
apastron and periastron increases with the eccentricity of the
orbit. The skew angle $\mu$ can affect the proximity of regions of the
shock cap to the primary star around periastron
(Fig.~\ref{fig:shkcap_skew}), and the resulting level of occultation
and attenuation.

\subsection{The ballistic CD}
\label{subsec:bcd}

To construct the large-scale 3D structure of the WCR, gas packets are
released from the endpoints of the shock cap at specific phase
intervals with a velocity equal to the slower wind, $v_{\rm sl}$.
The $x, y$, and $z$ components of the velocity of gas leaving the end
of the shock cap at a specific orbital phase are given by:

\begin{figure}
\begin{center}
\psfig{figure=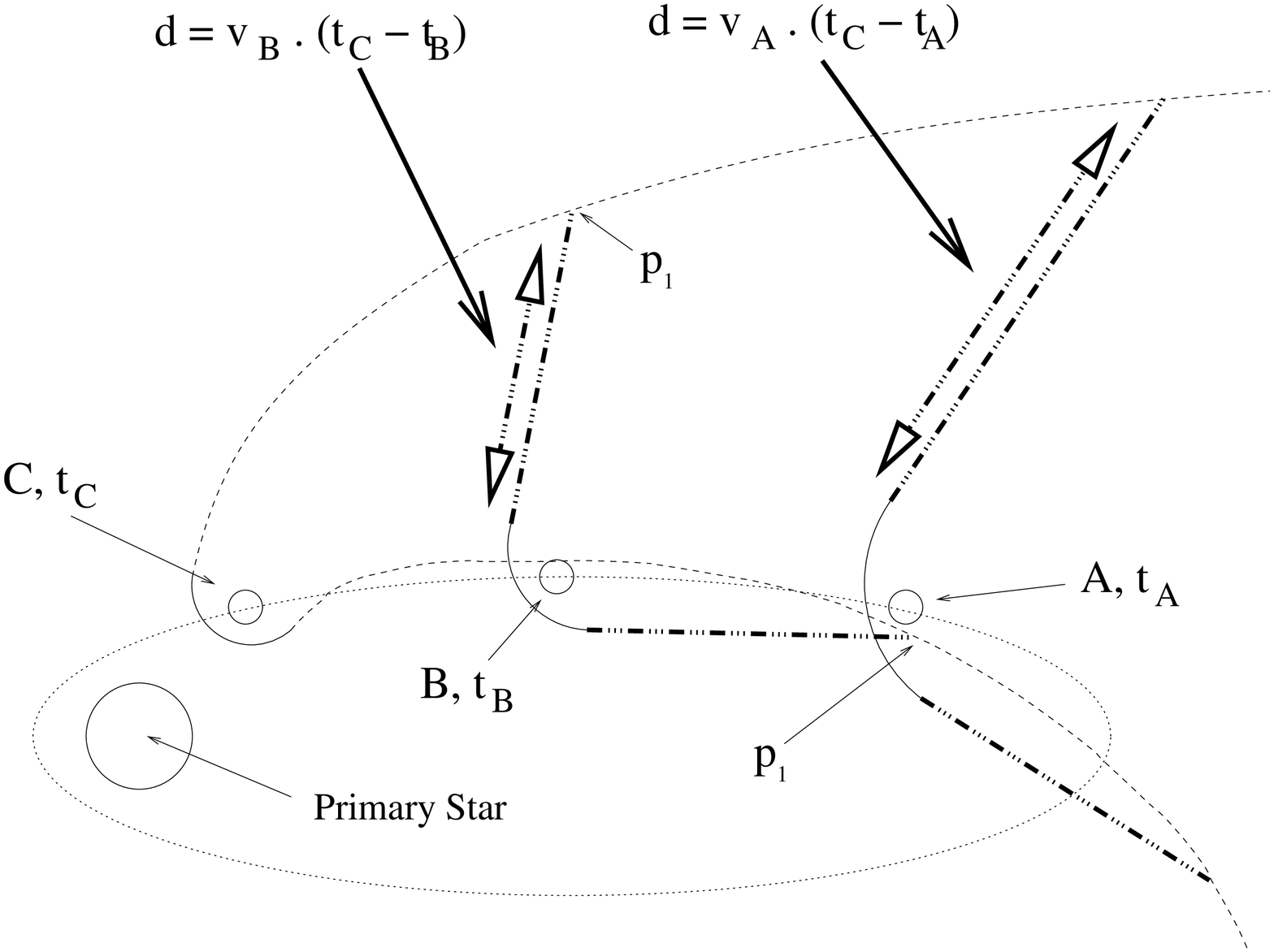,width=8.0cm}
\end{center}
\caption[]{Schematic diagram showing how position vectors to points on
the ballistic CD are obtained. If the secondary star is currently at
position C, points p$_{1}$ on the ballistic CD are found by
determining the position and flow direction of gas leaving the end of
the shock cap when the star is at position B and advecting the flow by
the time difference between these two orbital phases. This process is
repeated until the position of the ballistic CD has been traced back
the desired number of orbits (in this work the structure is traced
back over 2 orbits). Distances of 100's-1000's of au can easily be
covered, as required.}
\label{fig:cdcoords_howto}
\end{figure}

\begin{eqnarray*}
   \hat{v}_{x} = & v_{\rm sl}\cos(\omega -\mu + \lambda\cos\zeta),\\
   \hat{v}_{y} = & v_{\rm sl}\sin(\omega -\mu + \lambda\cos\zeta),\\
   \hat{v}_{z} = & v_{\rm sl}\sin\zeta \\
\end{eqnarray*}
where $\lambda$ = $\theta_{\rm c1\infty}$ is the asymptotic
half-opening angle of the contact discontinuity viewed from the star with
the stronger wind.

The ballistic part of the CD is then constructed by considering a
sequence of previous positions of the ballistic points at the
termination of the shock cap, and the current position of the gas flow
from these points given that they move along linear trajectories (see
Fig.~\ref{fig:cdcoords_howto}).

The position of points on the ballistic part of the CD ($x_{\rm{CD}},
y_{\rm{CD}}$, and $z_{\rm{CD}}$) at the time $t$ is given by their
position at the time they were emitted from the end of the shock cap
($x_{\rm{cap}}, y_{\rm{cap}}$, and $z_{\rm{cap}}$) plus the distance
they have since travelled at velocity
$\hat{v}_{\rm{x}},\hat{v}_{\rm{y}},\hat{v}_{\rm{z}}$, i.e.

\begin{eqnarray*}
   x_{\rm{CD}}(t) = & x_{\rm{cap}}(t-T) + \hat{v}_{\rm{x}}(t-T)T,\\
   y_{\rm{CD}}(t) = & y_{\rm{cap}}(t-T) + \hat{v}_{\rm{y}}(t-T)T,\\
   z_{\rm{CD}}(t) = & z_{\rm{cap}}(t-T) + \hat{v}_{\rm{z}}(t-T)T,
\end{eqnarray*}

\noindent where $T$ is the time elapsed since the flow left the
end of the shock cap.

\begin{figure}
\begin{center}
    \begin{tabular}{c}
      \resizebox{90mm}{!}{{\includegraphics{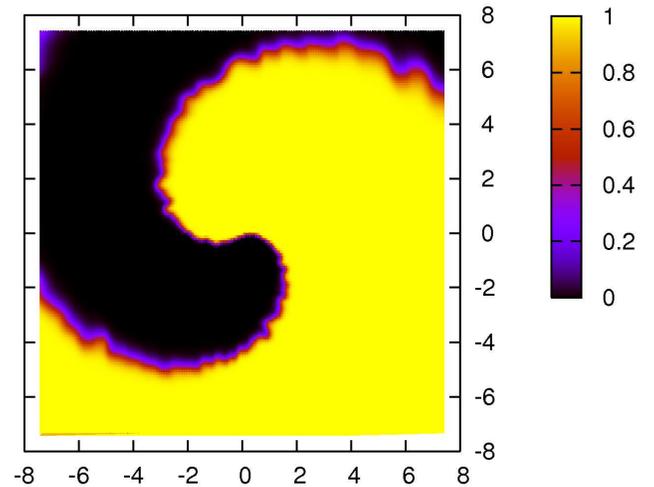}}} \\
\resizebox{100mm}{!}{{\includegraphics{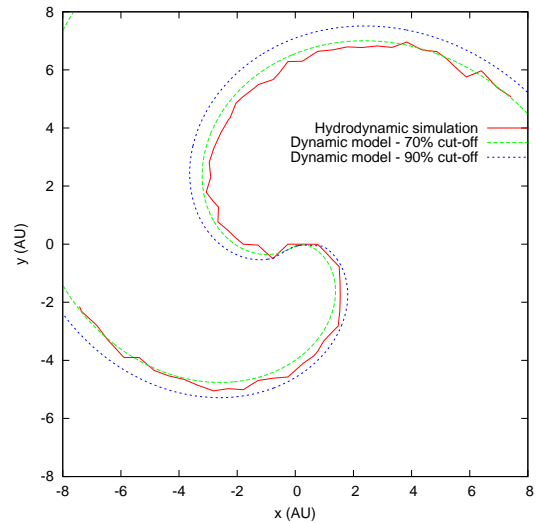}}} \\
    \end{tabular}
    \caption[]{3D plots viewed from above the orbital plane showing
    the position of the contact surface separating the stellar winds
    from a hydrodynamic simulation (top) and from the dynamical model
    with a varying cut-off percentage for the transition between the
    shock cap and the ballistic CD (bottom). The shading in the
    hydrodynamic simulation identifies the different stellar winds,
    where a value of 0.5 marks the CD. In these simulations an orbital
    period of 19.7 days, eccentricity, $e$ = 0.0, stellar masses,
    $M_{1} = M_{2} = 30 \Msol$, terminal wind speeds of $v_{\infty1} =
    v_{\infty2} = 1500\;\rm{kms}^{-1}$ were used. The mass-loss rates
    of the primary star was $\dot{M}_{1} = 3\times10^{-7}\Msolpyr$ and
    that of the secondary star was $\dot{M}_{1} =
    1\times10^{-7}\Msolpyr$.}
    \label{fig:comparison}
\end{center}
\end{figure}

\begin{figure*}
\begin{center}
    \begin{tabular}{cc}
      \resizebox{90mm}{!}{{\includegraphics{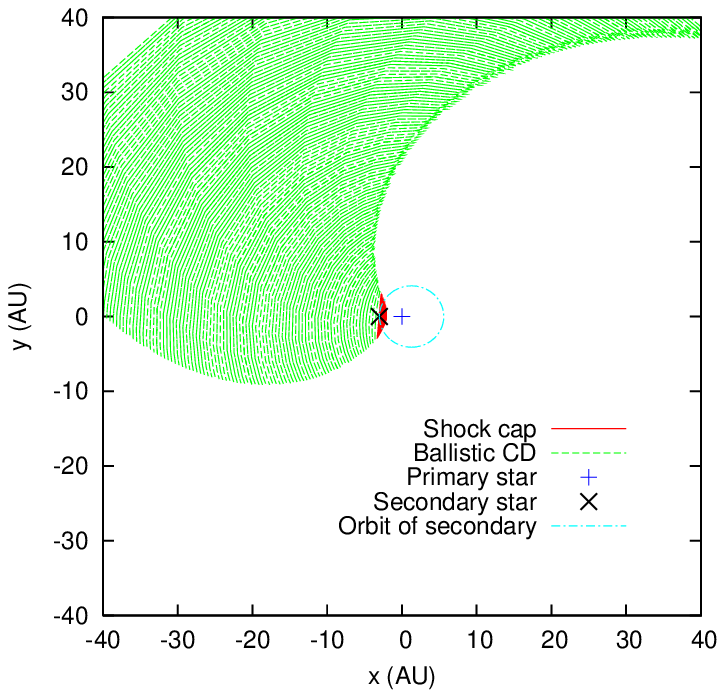}}} &
\resizebox{90mm}{!}{{\includegraphics{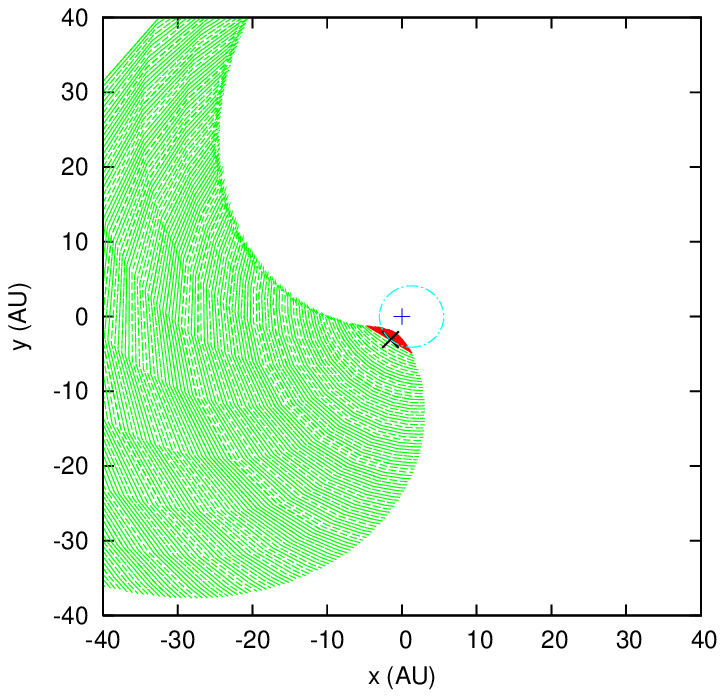}}} \\
      \resizebox{90mm}{!}{{\includegraphics{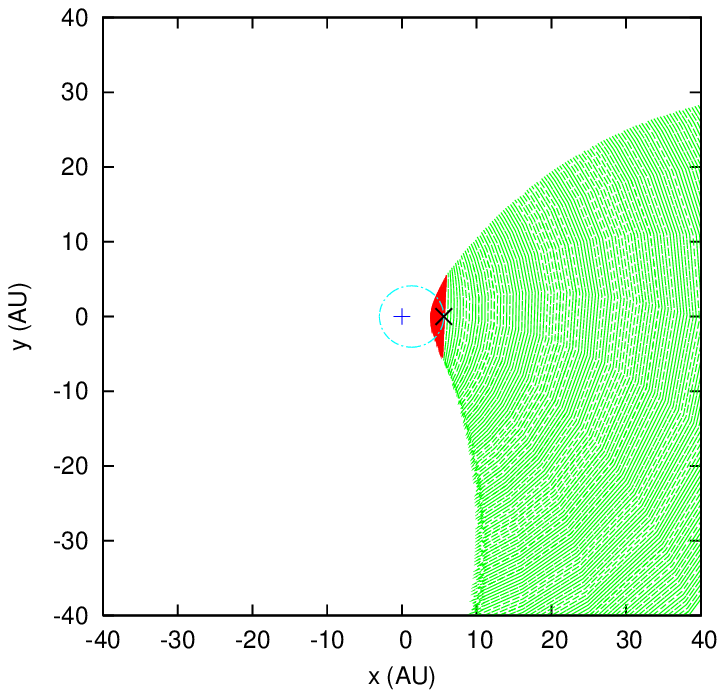}}} &
\resizebox{90mm}{!}{{\includegraphics{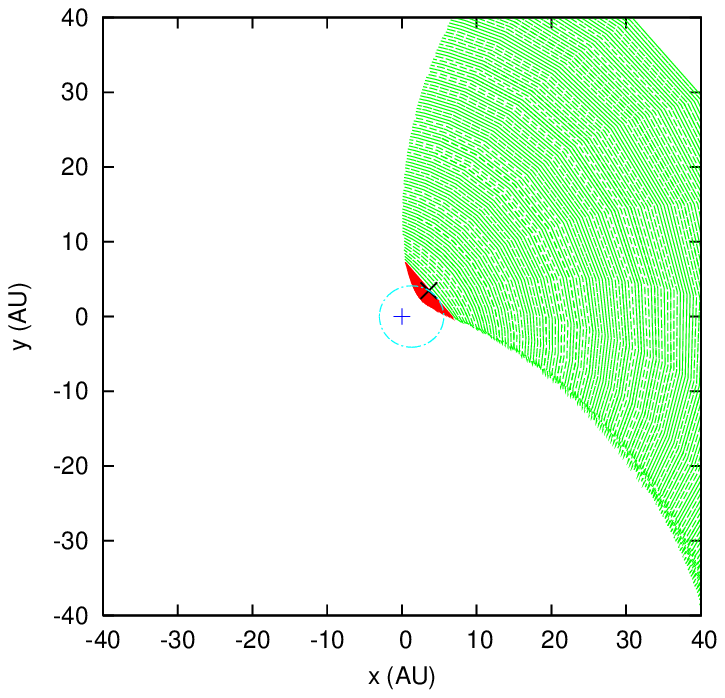}}} \\
    \end{tabular}
    \caption[]{3D plots viewed from above the orbital plane showing
    the position of the shock cap (red) and the structure of the
    ballistic CD (green) at various orbital phases from a simulation
    with $e = 0.3$ (see Table~\ref{tab:models} for the other relevant
    parameters). The skew to the shock cap due to the orbital motion
    of the secondary star is included. From top left to bottom right:
    $\phi = 0.0, 0.1, 0.5, \rm{and}\thinspace 0.7$. The curvature of
    the ballistic CD at $\phi = 0.0$ is caused by the high velocity of
    the secondary star around periastron passage. Note that the shocks
    either side of the CD are not displayed. The white area downstream
    of the ballistic CD in the corner of each plot is an artifact of
    the plotting software.}
    \label{fig:cdcoords_phi}
\end{center}
\end{figure*}

The number of coordinates in the ballistic CD is dependant on the
number of phase steps around the orbit, the number of orbital
revolutions followed, and the number of azimuthal steps
(i.e. $\delta\zeta$). In this work, the ballistic CD consists of 2000
coordinate positions along each azimuthal trajectory (1000 per orbit
traced). 

Tests performed using a 3D hydrodynamics code confirm that the
Coriolis force, which causes the curvature to the WCR, becomes
significant once the flow from the stagnation point is accelerated to
70\% - 90\% of the terminal speed of the slower wind
(Fig.~\ref{fig:comparison}), and the gas is at a distance from the
stars of order the stellar separation. Both of these conditions are
satisfied by the 85\% cut-off attained via calibration of the dynamic
model against hydrodynamic models. Interestingly, varying the value of
the cut-off percentage has the effect of improving the fit to one
spiral arm but reducing the quality of the fit to the other arm. Using
the Model A paramters (Table~\ref{tab:models}) the off-axis distance
of the ballistic point from the line of centres, $r_{\rm{max}}$,
increases by a factor of $\sim 3$ between 70 \% and 90 \%
(Table~\ref{tab:cutoffs}), whereas the opening angle of the shock
increases by roughly a half with a more linear relation. In
\S~\ref{sec:results} we show that there is little difference in the
X-ray lightcurves when this percentage is varied slightly.

\begin{table}
\begin{center}
\caption[]{Transition points between the shock cap and the ballistic
CD for varying percentages of the postshock flow
velocity. $r_{\rm{max}}$ is calculated using
Eq~\ref{eqn:zandr2}. $\theta_{\rm c1\infty}$ is the opening angle of
the shock at the transition point measured from the primary star (see
Fig.~\ref{fig:cantofig1}).}
\begin{tabular}{lll}
\hline
Cutoff & $r_{\rm{max}}$ &  $\theta_{\rm c1\infty}$\\
$(\%)$ & $(d_{\rm{sep}})$ & $(^{\circ})$ \\
\hline
70 & 0.53 & 33 \\
80 & 0.78 & 40 \\
85 & 0.99 & 44 \\
90 & 1.40 & 49 \\
\hline
\label{tab:cutoffs}
\end{tabular}
\end{center}
\end{table}

Fig.~\ref{fig:cdcoords_phi} shows the effect of the motion of the
stars on the ballistic CD on scales of the order of the semi-major
axis. The curvature of the CD close to the end of the shock cap is
greatest when the relative orbital velocity of the stars is high.  The
smooth connection of the ballistic CD to the shock cap indicates that
the assumptions inherent in the model are good at this level.

\begin{figure*}
\begin{center} 
    \begin{tabular}{cc}
      \resizebox{90mm}{!}{{\includegraphics{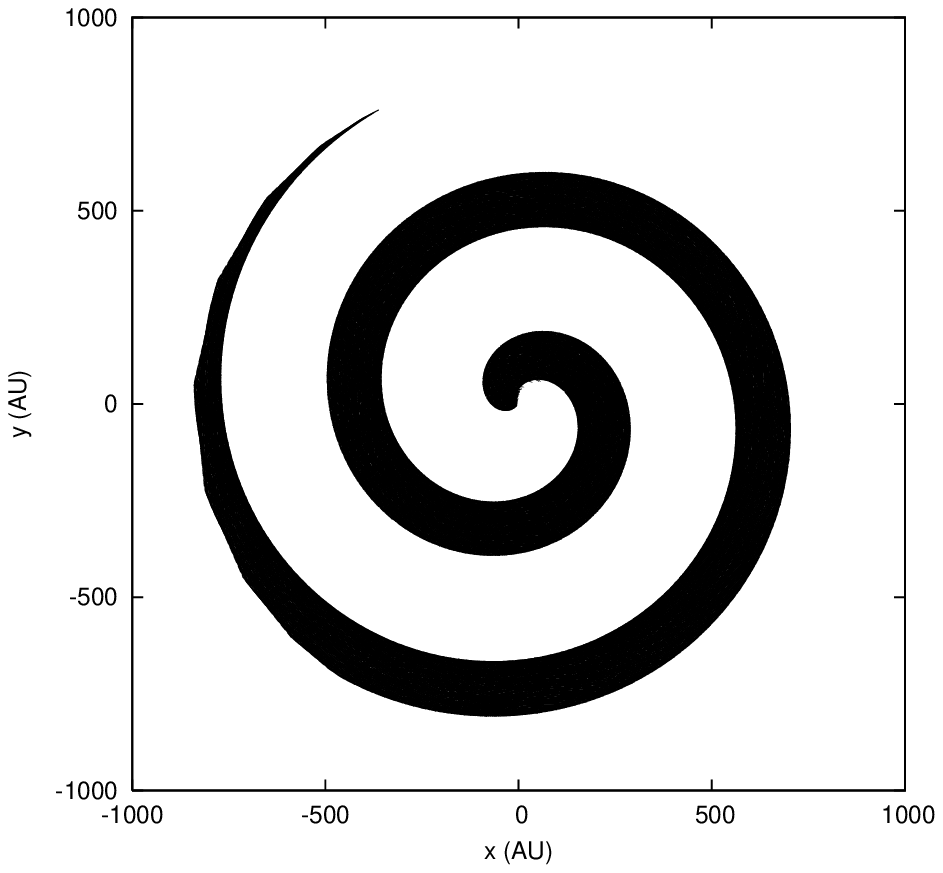}}} &
\resizebox{90mm}{!}{{\includegraphics{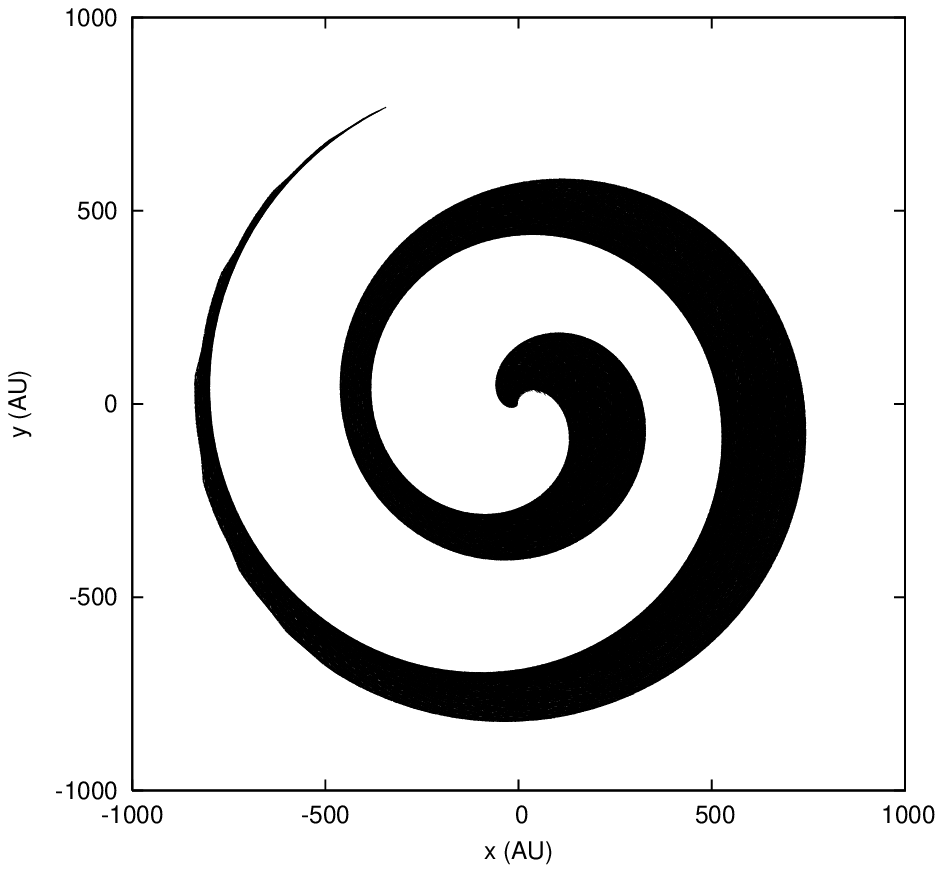}}} \\
      \resizebox{90mm}{!}{{\includegraphics{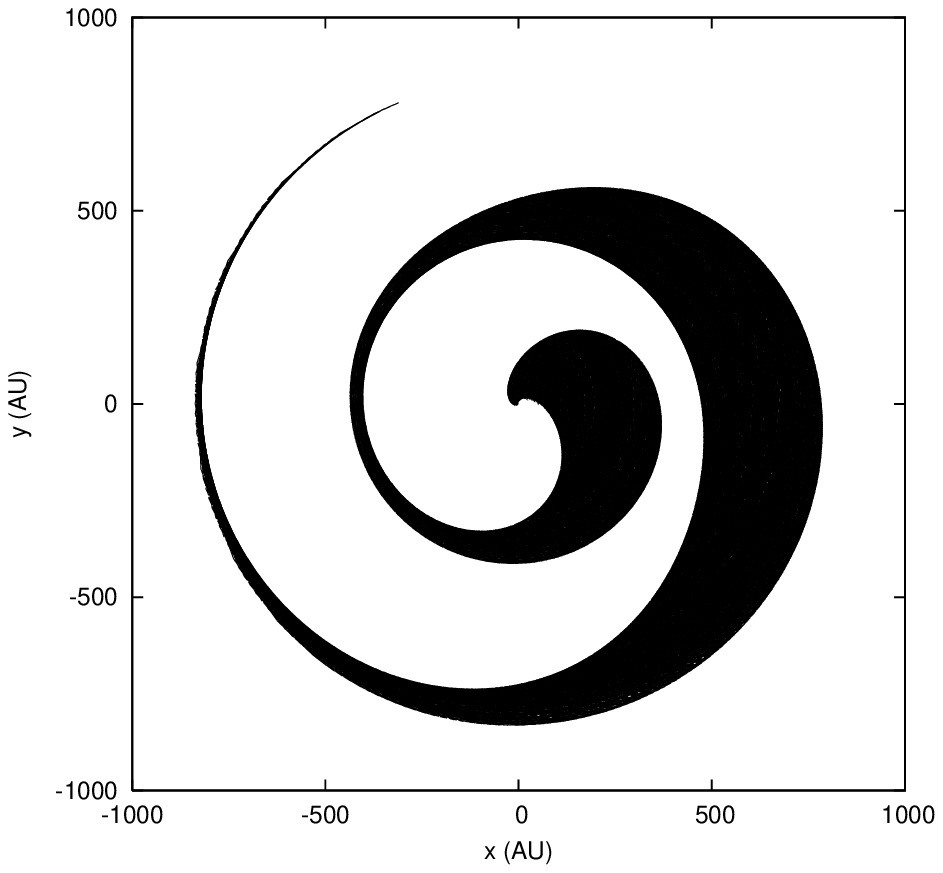}}} &
\resizebox{90mm}{!}{{\includegraphics{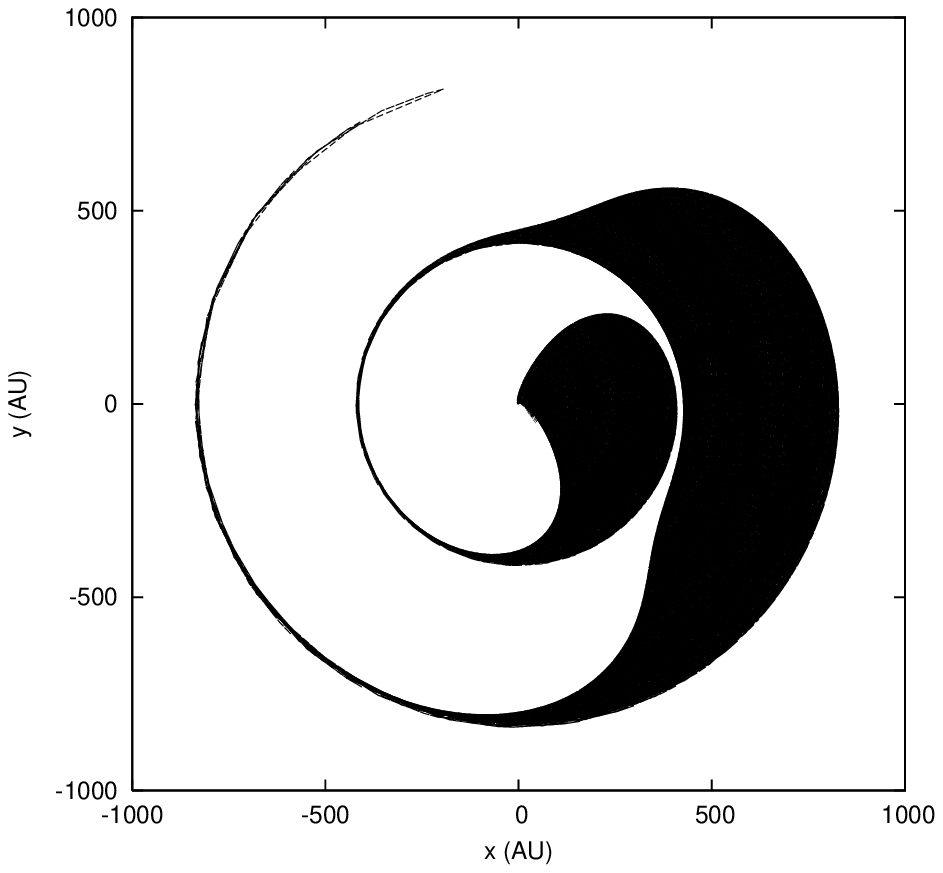}}} \\
    \end{tabular}
    \caption[]{Plots of the large-scale structure of the ballistic CD
    in the orbital plane at $\phi = 0.0$ as a function of
    eccentricity. $e = 0.0, 0.3, 0.6, \rm{and}\thinspace 0.9$ from top
    left to bottom right, with the other parameters given in
    Table~\ref{tab:models}. The semi-major axis of the orbit
    is $4.30\;$au, and the orbital period is $1\;$yr. The black region
    on the plots shows the projection of the 3D contact discontinuity
    onto the orbital plane, which encompasses the unshocked secondary
    wind. The white region mostly traces the unshocked primary wind.
    Note that the shocks either side of the CD are not displayed.  The
    variation of the velocity and separation of the stars around the
    orbit increases with the orbital eccentricity, which results in
    the increasing asymmetry of the projected CD. The tapering of the
    black region at the tail of the spiral (most noticeable in the $e
    = 0.0$ and $e = 0.3$ plots) is due to the fact that the ballistic
    part is only traced back for two orbits and that the leading and
    trailing arms of the CD finish in different directions. Further
    details of the interaction at $e = 0.3$ are shown in
    Fig.~\ref{fig:cdcoords_phi}. }
    \label{fig:cdcoords_ecc}
\end{center}
\end{figure*}

The structure of the ballistic CD at large scales is shown in
Fig.~\ref{fig:cdcoords_ecc} for a range of orbital eccentricities.  At
low orbital eccentricities, the spiral structures resemble the 3D
hydrodynamical models of \cite{Walder:2000,Walder:2002} and
\cite*{Lemaster:2007}, the dust spiral models of the pinwheel nebula
WR\thinspace104 by \cite{Harries:2004} and \cite{Tuthill:2008}, and
the CWo model for symbiotics developed by \cite{Kenny:2007}. Note,
however, that this figure shows the projection of the CD onto the
orbital plane, and not the position of the shocks either side of
it. If the shocked region were largely adiabatic, the shocks would
stand off from the CD and the width of the spiral structure on the
orbital plane would be somewhat greater.

At $e=0.9$ the secondary star moves very quickly through periastron,
resulting in the projected CD (which encompasses the region of
unshocked secondary wind) thinning to the left of the stars. In
contrast, there exists a large region of unshocked secondary wind to
the right of the stars, as the secondary star moves slowly around
apastron. This creates a low density cavity in the primary wind. The
X-ray attenuation in such systems will depend on the orbital phase, as
well as being sensitive to the position of the observer, and in
principle may vary widely. For instance, in a system like $\eta\;$Car,
the primary wind is very dense and much more strongly absorbing than
the secondary wind. An observer located on the positive $x$-axis at
infinity will predominantly view through the low density unshocked
wind of the secondary star, whereas an observer on the negative
$x$-axis will predominantly view through the high density unshocked
wind of the primary star. As the column density scales directly with
the density of the gas, these observers will see significantly
different X-ray lightcurves. On the other hand, if the primary wind is
more rarefied than the secondary wind, this behaviour reverses.

Finally, we note that in systems with highly eccentric orbits, the
amount of attenuation at phases around apastron may depend on the skew
angle of the shock cap which occurs around periastron.  This is
because the skew angle of the shock cap affects the duration and phase
where primary/secondary wind material is emitted in a certain
direction.  Depending on the viewing angle into the system, the
inclusion of aberration effects may result in a variation in the
attenuation to emission concentrated near the apex of the shock cap
due to the alteration in path length through the more strongly
absorbing wind.

\begin{table*}
\begin{center}
\begin{tabular}{lllllllllll}
\hline
Model & $M_{1}$ & $M_{2}$ & $\dot{M}_{1}$ & $\dot{M}_{2}$ & $\eta$ & $\theta_{\rm c1\infty}$ & $P$ & $a$ & $\chi_{1}$ & $\chi_{2}$ \\
 & (\Msol) & (\Msol) & (\Msolpyr) & (\Msolpyr) & & ($^{\circ}$) & & (au) &  &  \\
\hline
A & 50 & 30 & $1.0\times10^{-6}$ & $2.0\times10^{-7}$ & 0.20 & 62.6 & 1 yr & 4.3 & 100 & 500 \\ 
B & 50 & 30 & $1.0\times10^{-6}$ & $2.0\times10^{-7}$ & 0.20 & 62.6 & 1 month & 0.81 & 20 & 100 \\
C & 50 & 50 & $2.0\times10^{-5}$ & $1.0\times10^{-6}$ & 0.05 & 41.0 & 1 yr & 4.3 & 5 & 100 \\
\hline
\end{tabular}
\caption[]{Summary of the wind and orbital parameters of the model
systems. $\eta $ is the wind momentum ratio, and $a$ the semi-major
axis of the orbit.  In each model the wind speeds and stellar radii
adopted were $v_{\infty1} = v_{\infty2} = 2000\;\rm{kms}^{-1}$ and
$R_{\ast1} = R_{\ast2} = 10 \Rsol$. Cooling parameters are calculated
for an orbital separation of $d_{\rm{sep}} = a$. The half opening
angles, $\theta_{\rm c1\infty}$, of the WCR are comparable to the
$61^{\circ}$ and $41^{\circ}$ for Models A and C respectively
calculated using Eq.(3) of \cite{Eichler:1993}.}
\label{tab:models}
\end{center}
\end{table*}

\section{An example application - X-ray emission and absorption in an 
early-type binary}
\label{subsec:emission}
As an example application of the model described in
\S~\ref{sec:shkconemodel}, we consider the X-ray emission from
hypothetical O+O and WR+O-star colliding wind binaries. Model A is an
O+O binary with an orbital period of $1\;$yr and semi-major axis
$a=4.3\;$au. Model B examines the increasing effects of absorption as
the orbital period is reduced to $1\;$month. The third model system
(Model C) consists of a WR star with a mass-loss rate of
$2\times10^{-5}\Msolpyr$. The high velocities of the stellar winds are
sufficient to cause the postshock gas to emit at X-ray wavelengths,
and both winds are essentially adiabatic in the systems considered
(see Table~\ref{tab:models}). For the three models considered we use a
distance of 1 kpc, ISM column of $5\times10^{21}\;\rm{cm}^{-2}$, and
orbital eccentricity, $e=0.3$.

\subsection{The X-ray emission}
The X-ray emission from the WCR is a function of the gas temperature
and density.  Since the dynamical model discussed in the previous
section does not contain such information, we use a grid-based, 2D
hydrodynamical calculation of an axis-symmetric WCR to obtain
this. The numerical code is second-order accurate in time and space
\citep{Falle:1996,Falle:1998}. The resulting emission as a function of
off-axis distance is then mapped onto the coordinate positions in the
3D dynamical model. In this way we obtain the benefit of effectively
modelling the thermodynamic and hydrodynamic behaviour responsible for
the production of the X-ray emission, while simultaneously accounting
for the effect of the motion of the stars on the large-scale structure
of the WCR and the subsequent wind attenuation. Since the hydrodynamic
calculation is 2D, the computational requirements remain low.

\begin{figure}
\begin{center}
    \begin{tabular}{l}
      \resizebox{80mm}{!}{{\includegraphics{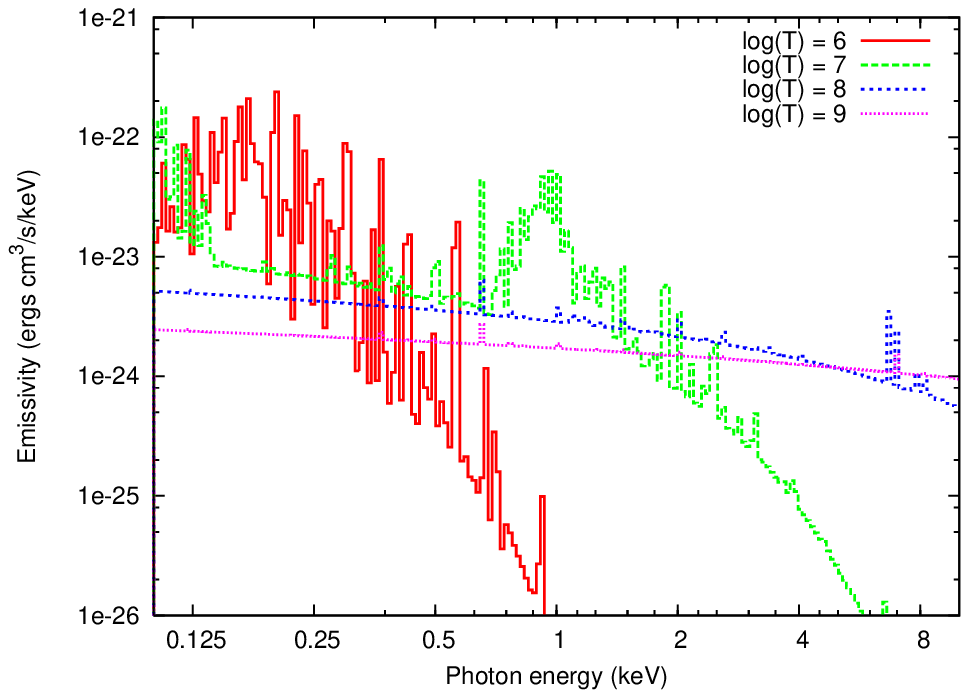}}} \\
    \end{tabular} 
\caption[]{Emissivity of the solar abundance gas in the spectral
  energy range 0.1-10.0 keV at various temperatures (K) as calculated
  from the \textit{MEKAL} thermal plasma code.}
\label{fig:emiss_spec}
\end{center}
\end{figure}

\begin{figure}
\begin{center}
    \begin{tabular}{l}
      \resizebox{80mm}{!}{{\includegraphics{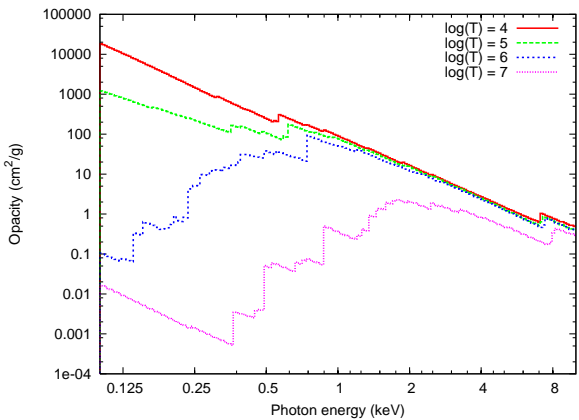}}} \\
    \end{tabular} 
\caption[]{Opacity (gcm$^{-2}$) in the spectral energy range 0.1-10.0
  keV at various temperatures (K).}
\label{fig:abs_spec}
\end{center}
\end{figure}

The X-ray emission calculated from each hydrodynamic cell in the WCR
is $\Gamma(E)=n^{2}V\Lambda(E,T)$, where $n$ is the gas number density
($\cm^{-3}$), $V$ is the cell volume ($\cm^{3}$), and $\Lambda(E,T)$
is the emissivity as a function of energy $E$ and temperature $T$ for
optically thin gas in collisional ionization equilibrium ($\rm{erg\;
cm}^{3}s^{-1}$). $\Lambda(E,T)$ is obtained from look-up tables
calculated from the \textit{MEKAL} plasma code \citep[][and references
therein]{Leidahl:1995} containing 200 logarithmically spaced energy
bins in the range 0.1-10.0 keV, and 101 logarithmically spaced
temperatures from $10^{4}$ to $10^{9} \;\rm{K}$. Solar abundances are
assumed for the O-star winds and the WR wind is assumed to have WN8
abundances (mass fractions of: H/He=0, C/He = $1.7\times10^{-4}$, N/He
= $5\times10^{-3}$, and O/He = $1\times10^{-4}$). The emissivity of
solar abundance gas is shown in Fig.~\ref{fig:emiss_spec} and the
corresponding opacity is shown in Fig.~\ref{fig:abs_spec}. The WN8
emissivities are very similar to those at solar abundance. Opacity
values are also similar for solar and WN8 abundances, with the most
significant difference being a factor of 2 increase at $10^{4}\;$K at
energies below $\sim 1\:$ keV.

\begin{figure}
\begin{center}
\psfig{figure=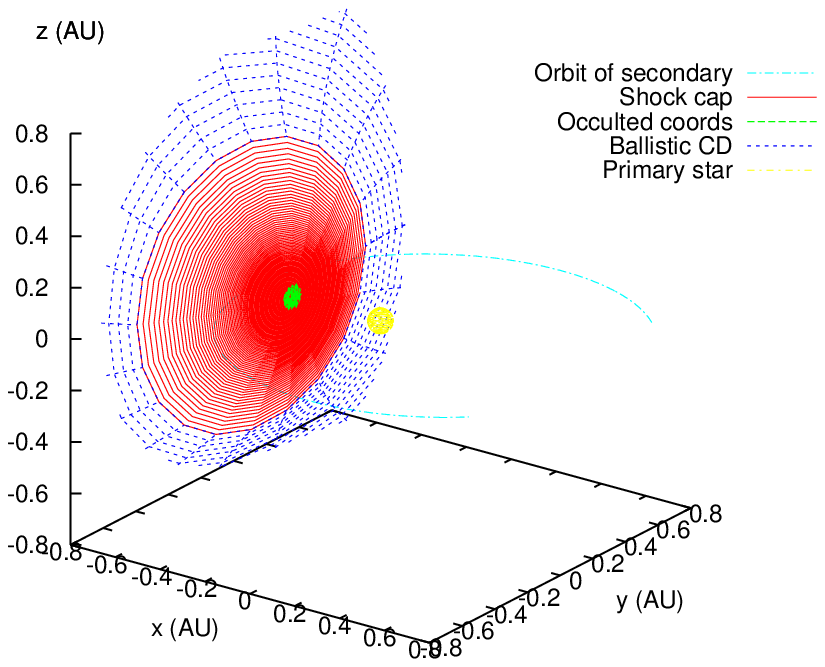,width=9.0cm}
\psfig{figure=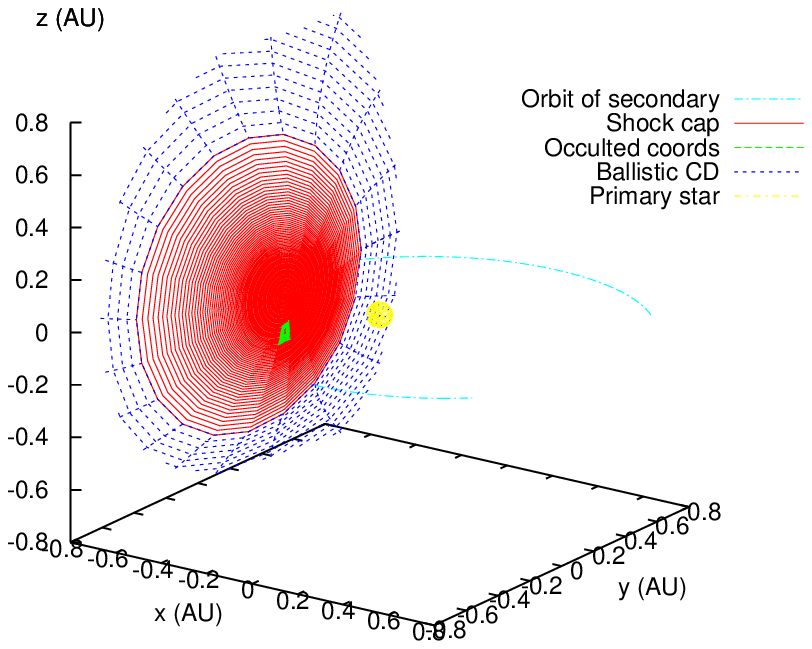,width=9.0cm}
\caption[]{3D plots showing occultation effects for two different
viewing angles at periastron ($\phi=0.0$) with $e = 0.3$ and an
orbital period $P = 1\;$month. The occulted emission from the shock
cap is represented in green (this is essentially the ``shadow'' of the
occulting star - the secondary star is hidden behind the shock
cap). The top panel shows the case where $i = 90^{\circ}$ and
$\theta=0^{\circ}$. The primary star is directly in front of the
secondary star and the apex of the WCR is occulted. In the bottom
panel $i = 70^{\circ}$ and $\theta=0^{\circ}$. The line-of-sight peers
over the top of the primary star and the ``shadow'' moves below the
apex of the WCR - less emission is occulted as a result.  The finite
number of coordinates on the shock cap, together with the
trapezium-shaped emission regions, account for the particular shape of
the ``shadow'' in this case. Aberration of the WCR is included in both
plots.}
\label{fig:occ_shkcaps}
\end{center}
\end{figure}

The emission values are then appropriately scaled for the changing
stellar separation around the orbit \citep[$L_{\rm{X}} \propto d_{\rm
sep}^{-1}$ in the adiabatic limit,][]{Stevens:1992} and placed onto
the 3D shock cap and ballistic CD. Emission values are assigned to
points within $3\;d_{\rm sep}$ of the apex of the WCR. This accounts
for $\sim 90$ per cent of the 0.1-10 keV emission and $> 99$ per cent
of the 2-10 keV emission.

\subsection{The attenuation}
\label{sec:absorption}

To compute X-ray lightcurves, the orientation of the observer relative
to the system must be specified. Since the model assumes the orbit of
the stars is in the $xy$ plane, viewing angles into the system can be
described by the inclination angle that the line-of-sight makes with
the $z$ axis, $i$, and the angle the projected line-of-sight makes
with the major axis of the orbit, $\theta$. Positive values of
$\theta$ correspond to projected lines of sight in the prograde
direction from the positive $x$ axis. The components of the unit
vector along the line-of-sight, $\underline{\hat{u}}$, are thus

\begin{eqnarray*}
   u_{x} & = & \cos\theta\sin i, \\
   u_{y} & = & \sin\theta\sin i, \\
   u_{z} & = & \cos i.
\label{eqn:losunitvectors}
\end{eqnarray*}

There are 3 ways in which the intrinsic X-ray emission can be attenuated.
First, it can be occulted by the stars (this effect, of course, is greatest 
in short period systems). Second, there will be some absorption through 
the un-shocked stellar winds. Finally, there will be attenuation through 
the shocked gas in the WCR. The latter is only important in systems 
where the shocked gas of at least one of the winds is strongly radiative,
otherwise the gas in the WCR remains hot and the attenuation through it
is small. But if there is significant cooling, as for example occurs when
the cool wind in symbiotic systems is shocked, a thin,
dense, and cold layer of gas is formed at the CD, which will be a
significant source of attenuation in the system. Significant cooling
of the WCR can also occur in early-type binary systems of which  
\etacar is an example. We now describe how attenuation by each of the 
above-mentioned methods is calculated in our model.

\begin{figure*}
\begin{center}
    \begin{tabular}{ll}
      \resizebox{80mm}{!}{{\includegraphics{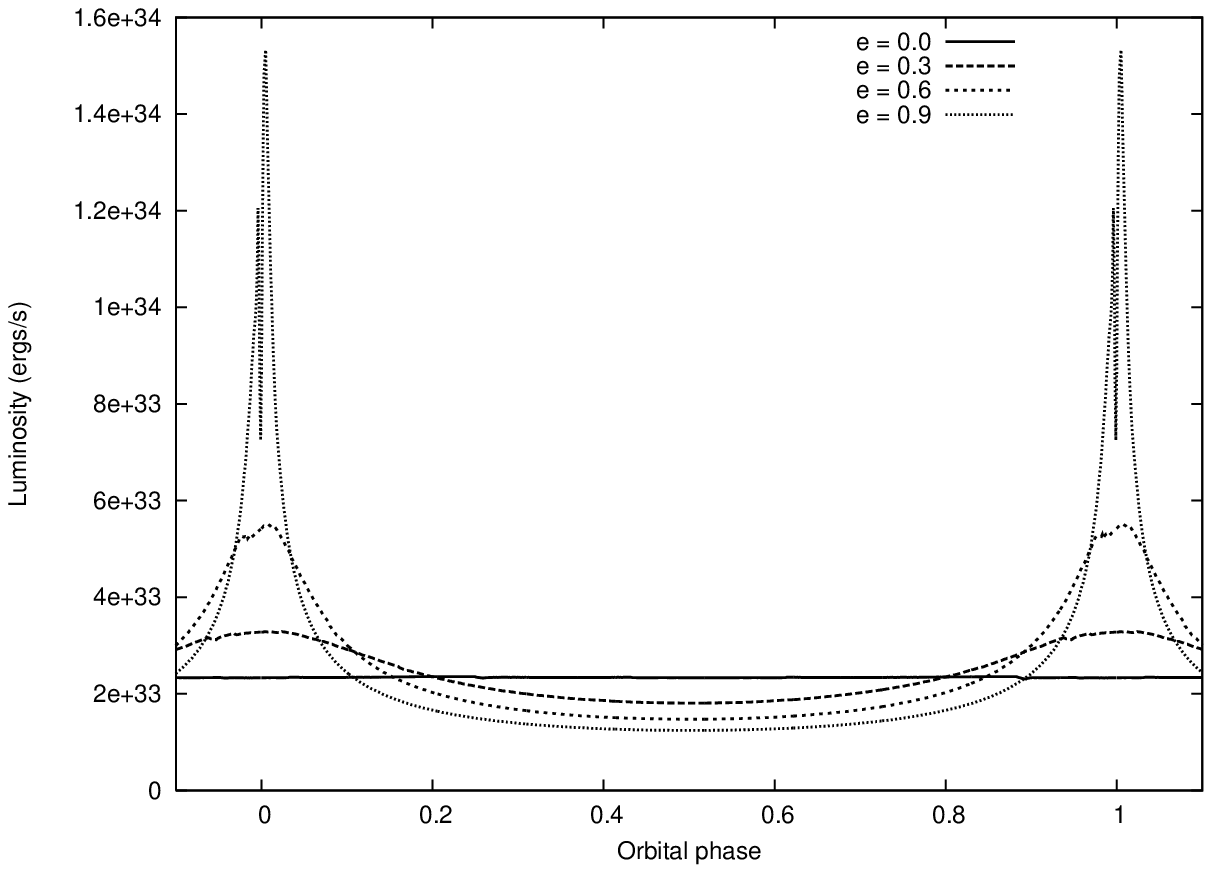}}} &
\resizebox{80mm}{!}{{\includegraphics{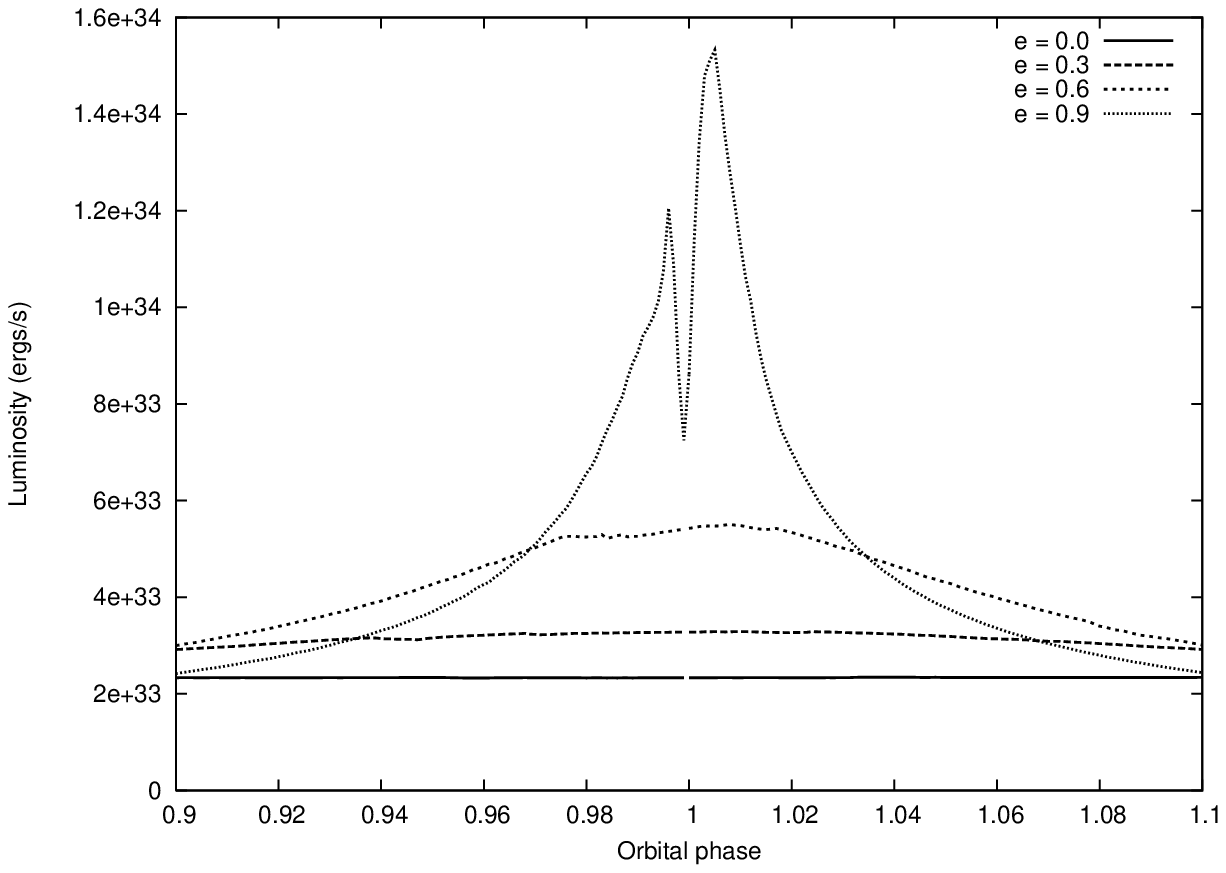}}} \\
    \end{tabular} 
\caption[]{0.1-10.0 keV lightcurves for Model B with $e = 0.0, 0.3,
  0.6, \rm{and}\thinspace 0.9$ with occultation included. Left: The
  variability shown over an entire orbital period. Right: The
  variation over the orbital phase range 0.9-1.1. Occultation is
  highest at periastron for the $e = 0.9$ lightcurve because the size
  of the shock cap relative to the primary star is smallest at this
  time. Orbital motion induced skew causes the pre-minimum intrinsic
  luminosity to be lower than the post-minimum, this effect increases
  with orbital eccentricity. For all curves $i=90^{\circ}$ and $\theta
  = 0^{\circ}$. Interstellar absorption has not been considered in
  these calculations.}
\label{fig:int_lc}
\end{center}
\end{figure*}

\subsubsection{Occultation by the stars}
\label{subsec:occultation}
An important line-of-sight effect in binary star systems is
occultation, particularly in the case of eclipsing binaries.  To
calculate this effect in our model, a line-of-sight is traced from
each emitting region on the shock cap and ballistic CD, and its
distance of closest approach to the centre of each star is calculated.
If this distance is less than the radius of the star, and the star is
in front of the emitting region, then occultation occurs, and none of
the emission from the emitting region being considered reaches the
observer.

A visual representation of the occultation of the WCR by the
primary star is shown in Fig.~\ref{fig:occ_shkcaps}. The degree of
occultation can be reduced by reducing the inclination angle
$i$ (since the strongest X-ray emission occurs at the apex of the WCR). 
Although not shown, the phase at which the maximum occultation
occurs can be altered by changing the value of $\theta$.

Occultation causes little change to the observed luminosity over the
entire orbit for the Model A system. This is due to the relatively
small size of the stars in comparison to the extended emitting region
for the 0.1-10.0 keV X-rays. Occultation effects become more
noticeable in shorter period systems, and/or those with highly
eccentric orbits (since the linear size of the shock cap is $\propto
1/d_{\rm{sep}}$).  Occultation is also favoured where one (or both) of
the stars has a large stellar radius (e.g. $\eta\;$Car, Parkin et al.,
in preparation), and when $i$ is large. For instance, the $e=0.9$
lightcurve in Fig.~\ref{fig:int_lc} shows a pronounced occultation
effect at orbital phase $\phi \simeq 1.00$, during which the emission
falls sharply by a factor of 2. The width of the minimum due to
occultation effects is very narrow as the high eccentricity means that
the stars move rapidly through periastron, but the depth of the
minimum is large ($\sim 75$\% of the intrinsic 2-10 keV emission is
occulted).

\subsubsection{Absorption by the un-shocked stellar winds}
\label{subsec:windattenuation}

For inclinations, $| i | \geq \pi/2 - \theta_{\rm c1\infty}$, the
line-of-sight from emitting regions near the apex of the WCR will
intersect the CD numerous times as it spirals out, and thus traverses
first through one wind and then the other, etc. The total column
density along a line-of-sight is then the sum of the individual column
densities along the specific distances travelled in each
wind. Accurate knowledge of where the line-of-sight intersects the CD,
and the density of the gas at any point in space is therefore required
if the total column density along a given sight line is to be
calculated.

To determine if and where an intersection through the CD occurs, the
shock cap and ballistic CD are tesselated into a sequence of
triangular planar facets constructed between three neighbouring
coordinates ($\underline{P}_{\rm{a}}$, $\underline{P}_{\rm{b}}$, and
$\underline{P}_{\rm{c}}$). To determine if the line-of-sight
intersects a given triangle the normal to the plane in which the
triangle lies, $\underline{\hat{n}}$, is calculated from

\begin{equation}
   \underline{\hat{n}} =
   (\underline{P}_{\rm{b}}-\underline{P}_{\rm{a}})\times(\underline{P}_{\rm{c}}-\underline{P}_{\rm{a}}).
\label{eqn:intnormal}
\end{equation}

\noindent The dot product of $\underline{\hat{n}}$ with the
line-of-sight vector gives the angle between the line-of-sight and the
plane. If the resultant angle is non-zero the line-of-sight vector
will intersect the plane in which the triangular facet lies at some
point in space.

The component vectors to the intersection point ($x
_{\inter}$, $y _{\inter}$, and $z _{\inter}$)
are found by substituting the line parameter at the intersection
point,

\begin{equation}
   \kappa = \frac{n_{\rm{x}}x_{\cap} +
   n_{\rm{y}}y_{\cap} +
   n_{\rm{z}}z_{\cap} +
   n_{\rm{const}}}{n_{\rm{x}}u_{\rm{x}} +
   n_{\rm{y}}u_{\rm{y}} +
   n_{\rm{z}}{u_{\rm{z}}}},
\label{eqn:intlinepar}
\end{equation}

\noindent into the equations

\begin{eqnarray*}
   x_{\inter} = & x_{\cap} + \kappa u _{\rm{x}},\\
   y_{\inter} = & y_{\cap} + \kappa u _{\rm{y}},\\
   z_{\inter} = & z_{\cap} + \kappa u _{\rm{z}},
\end{eqnarray*}

\noindent where the equation of the plane with normal
$\underline{\hat{n}}$ and vector components $n_{\rm{x}}$,
$n_{\rm{y}}$, and $n_{\rm{z}}$ is

\begin{equation}
   n_{\rm{x}} \underline{x} + n_{\rm{y}}
   \underline{y} + n_{\rm{z}} \underline{z} +
   n_{\rm{const}} = 0
\label{eqn:plane}
\end{equation}

\begin{figure}
\begin{center}
\psfig{figure=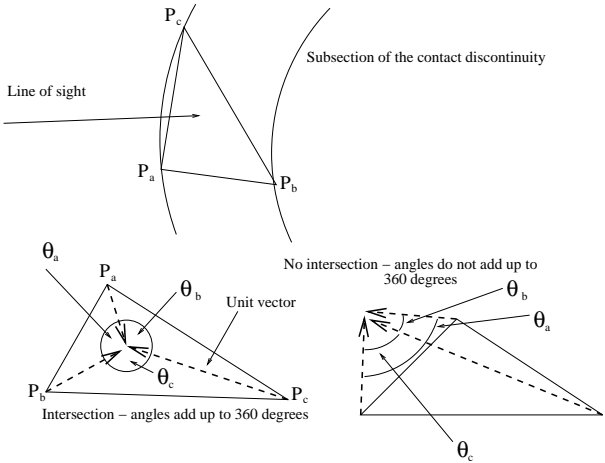,width=8.0cm}
\caption[]{The method used to determine whether a given line-of-sight
intersects the CD through a specific triangular tessel. Triangles are
constructed over the entire shock cap and ballistic CD from adjacent
coordinates and the unit vectors between the corners of the triangle
and the intersection coordinate are used to determine if the
intersection occurs within the boundary of the triangle.}
\label{fig:triangles}
\end{center}
\end{figure}

In general, the intersection occurs outside of the triangular
facet. Unit vectors are constructed between the corner points of the
facet and the intersection point to determine whether the intersection
occurs within its boundaries. The three dot products between these
three unit vectors gives the angles $\theta_{a}$, $\theta_{b}$, and
$\theta_{c}$. Only if the intersection point lies within the
boundaries of the triangular facet will the equation
$\theta_{a}+\theta_{b}+\theta_{c}=2\pi $ be satisfied (see
Fig.~\ref{fig:triangles}). By looping over the entire sequence of
triangles, every possible intersection of the line-of-sight with the
CD is determined.

With the coordinates of the intersection points
($x_{\inter}$, $y_{\inter}$, and $z_{\inter}$), it is a simple task to
calculate the column density through the unshocked winds,
$\sigma_{\rm{w}}$. Lines-of-sight which pass very close to the stars
sample the acceleration region of the wind. Therefore, 
we use a $\beta$-velocity law of the form

\begin{equation}
   v(r) = v_{\infty}\left(1 -
   \frac{R_{\ast}}{r}\right)^{\beta}
\label{eqn:betavellaw}
\end{equation}

\noindent to determine the density of the wind at radius $r$ from the
star.  $\beta$ describes the acceleration of the wind with $\beta=0.8$
appropriate for O star winds \citep{Lamers:1999}.  Because the width
of the WCR is not considered in our model, the volume of unshocked
wind and the resulting attenuation are overestimated, though this
approximation will not have a signifcant impact on our results.

\subsubsection{Absorption by the shocked stellar winds}
\label{subsec:sdattenuation}
As already mentioned, the attenuation of X-rays through the shocked
wind(s) needs to be considered if one or both winds strongly cool. In
the O+O and WR+O-star binaries considered in this section, the shocked
gas is largely adiabatic. However, for completeness, we discuss here a
methodology for calculating the absorption due to X-rays intersecting
a cold dense layer of postshock gas at the CD.  This is applied to
models of \etacar in Parkin et al. (in preparation). In
Figs.~\ref{fig:cutoff_lc}, \ref{fig:inc_lc}, \ref{fig:los_lc_1month},
\ref{fig:inc_lc_wr}, \ref{fig:los_lc_wr}, \ref{fig:all_col} and
\ref{fig:specs} this effect does not need to be considered.

The surface density, $\sigma_{s}$, of the postshock gas along the CD,
when both winds have $\chi \ltsimm 1$, has been computed by
\citet{Girard:1987}, \citet{Canto:1996}, and \citet{Kenny:2005}. In
each of these works, turbulence in the postshock flow is assumed to
fully mix the material from both winds and the surface density
calculated is for shocked gas from both winds. Alternatively, if only
one of the winds is radiative (i.e. the other remains largely
adiabatic), or the postshock flow is assumed not to mix, then the
surface density can be calculated from considering conservation of
mass flux \citep[e.g.,][]{Antokhin:2004}. To calculate the surface
densities in Figs~\ref{fig:abs_lc} and \ref{fig:abs_col} we have used
Eq.(30) of \cite{Canto:1996}.

In our model, the ballistic part of the WCR is asymmetric due to
orbital motion. Since the pre-shock flow is practically tangential to
the CD at this point, we calculate the total surface density of the
postshock winds (which in this subsection are assumed to cool) in this
region by considering conservation of mass flux. The surface density
of the postshock gas close to the apex of the WCR varies by over an
order of magnitude between periastron and apastron when $e=0.9$.

Since the width of the cool dense layer of gas alongside the CD is not
infinitely thin, the degree of absorption through it depends on the
angle subtended between the line-of-sight and the normal to its (i.e.
the CD's) surface, $\gamma$. The column density intersected by the
line-of-sight is therefore

\begin{equation}
   \sigma'_{\rm{s}} = \frac{\sigma_{\rm{s}}}{\cos \gamma},
\label{eqn:mod_sigma}
\end{equation}

\noindent where $\sigma_{s}$ is the actual surface density of the
cooled layer. When the line-of-sight becomes closely tangential to the
CD, $\sigma'_{\rm{s}}$ can become large, even if $\sigma_{\rm{s}}$
itself is not particularly large.
 
The maximum value of $\sigma'_{\rm{s}}$ is constrained by the
curvature of the WCR and the finite path length through the shocked
gas. To determine the maximum path length requires knowledge of the
width of the cooled layer and its radius of curvature at the point of
interest on the CD. On the shock cap the density of the cooled
postshock region, $\rho_{\rm{ps}}$, can be determined by equating the
ram pressure of the preshock gas with the thermal pressure of the
postshock gas \citep{Kashi:2007},

\begin{equation}
   \rho_{\rm{ps}} = \frac{m_{\rm{H}} \rho_1(r) (v_{\infty1}
   \sin\xi)^2}{k_{\rm{B}} T_{\rm{ps}}},
\label{eqn:rhops}
\end{equation}

\noindent where $T_{\rm{ps}}$ is the temperature of the cooled
postshock gas ($T_{\rm{ps}}$ is taken to be $10^4\;$K), $\rho(r)$ is
the preshock gas density as a function of distance from the respective
star, $\xi$ is the angle between the preshock velocity vector and the
tangent to the shock surface, and $m_{\rm{H}}$ and $k_{\rm{B}}$ are
the mass of a hydrogen nucleus and Boltzmann's constant respectively.
The thickness of the cooled layer, $l_{\rm{shock}}$, is then

\begin{equation}
   l_{\rm{shock}} = \frac{\sigma_{\rm{s}}}{\rho_{\rm{ps}}}.
\end{equation}

\noindent The thickness of the cooled layer in the ballistic CD region
cannot be calculated in this manner because the shocks
are now fully oblique. Therefore, a linear extrapolation is used to
determine the downstream thickness.

The radius of curvature at a point on the shock cap is

\begin{equation}
   \Lambda = \left|\frac{ds}{d\hat{t}}\right|,
\end{equation}

\noindent where $ds$ is the distance between two points on the WCR and
$d\hat{t}$ is the difference in the unit vectors tangent to the WCR at
those two points. Consideration of the maximum path length through the
cool dense layer, $d_{\rm{max}}$, then gives the maximum value for
$\frac{\sigma'_{\rm{s}}}{\sigma_{\rm{s}}}$ as

\begin{equation}
   \left|\frac{\sigma'_{\rm{s}}}{\sigma_{\rm{s}}}\right|_{\rm{max}}
   \simeq \frac{d_{\rm{max}}}{l_{\rm{shock}}} =
   \sqrt{4+\frac{8\Lambda}{l_{\rm{shock}}}}.
\label{eqn:maxsigmamod}
\end{equation}

The skewing of the shock cap due to orbital motion will evoke an
asymmetry in the postshock gas density \citep{Lemaster:2007}. This is
naturally accounted for in Eqs.~\ref{eqn:rhops}
and~\ref{eqn:maxsigmamod}.

\subsubsection{The observed emission}
\label{subsec:observedemission}

\begin{figure}
\begin{center}
    \begin{tabular}{l}
      \resizebox{80mm}{!}{{\includegraphics{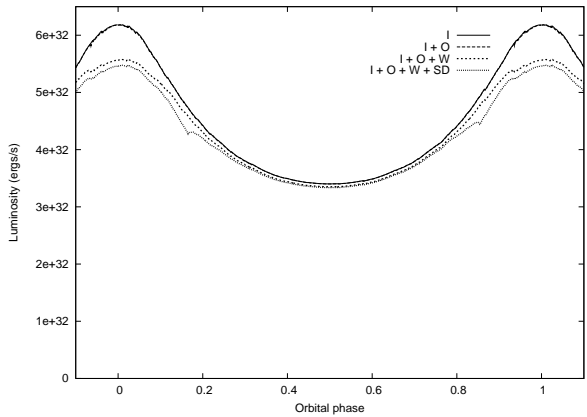}}} \\
    \end{tabular} 
\caption[]{The effect of the various attenuation mechanisms on the
observed 0.1-10.0 keV emission as a function of orbital phase for
Model A. The un-attenuated intrinsic emission (I), intrinsic emission
with occultation (I + O), intrinsic emission with occulatation and
wind attenuation (I + O + W), and intrinsic emission with occultation,
wind attenuation, and attenuation due to intersection of the shocked
gas (I + O + W + SD) are shown. Attenuation through the shocked gas
(SD) is only important if the shocked gas can cool efficiently and is
shown here purely for illustration of its effect. The assumed viewing
angles are $i = 90^{\circ}$ and $\theta = 0^{\circ}$, and the other
model parameters are listed in Table~\ref{tab:models}. The skew to the
WCR due to orbital motion is included. Eq.(30) of \cite{Canto:1996}
has been used to calculate surface densities as this provides an
approximate upper limit to the attenuation by shocked
gas. Interstellar absorption ($\sim 5\times10^{21}\rm{cm}^{-2}$) is
included. The variation in the intrinsic emission is proportional to
$1/d_{\rm{sep}}$.}
\label{fig:abs_lc}
\end{center}
\end{figure}

\begin{figure}
\begin{center}
    \begin{tabular}{l}
      \resizebox{80mm}{!}{{\includegraphics{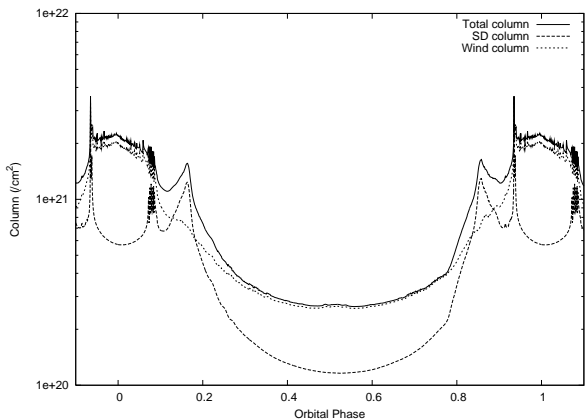}}} \\
    \end{tabular} 
\caption[]{Variation of emission weighted column density as a function
of orbital phase. Interstellar absorption is not included.}
\label{fig:abs_col}
\end{center}
\end{figure}

In the hypothetical binary systems considered in this paper, the
shocked gas in the WCR remains hot as it flows out of the system and
thus contributes insignificantly to the absorption. Hence the total
column density along a given line-of-sight is the sum of the column
densities through the unshocked winds. The attenuation declines as the
line-of-sight leaves the system, and is negligible at the distances
which our model extends to (the distance the wind flows over two
orbits).
 
Absorption cross-sections for solar abundance gas at $10^4\;$K are
used to obtain the optical depth, $\tau$, along specific
lines-of-sight in 200 logarithmically spaced bins over the energy
range 0.1-10.0 keV. The observed attenuated emission, $I_{\rm{obs}} =
I_{0} e^{-\tau}$, where $I_{0}$ is the intrinsic emission.

Fig.~\ref{fig:abs_lc} demonstrates the effect of including the various
attenuation mechanisms on the resultant emission. As previously
mentioned, occultation causes little reduction in emission because of
the minute size of the stars compared to the extended WCR
(Fig.~\ref{fig:occ_shkcaps}). For the assumed position of the
observer, absorption by the unshocked winds increases as the stars
approach each other and reaches a maximum at periastron. For
illustrative purposes we also show the attenuation that occurs if the
postshock gas cools and forms a thin dense layer along the CD (this
does not occur in the systems considered since the shocked gas remains
largely adiabatic as it flows out of the system). When the
line-of-sight becomes closely tangential to the WCR the path length of
X-rays through the shocked gas and the subsequent attenuation via this
mechanism reaches a maximum. This can be seen in the small dips in the
lightcurve at orbital phases 0.18 and 0.82, with corresponding peaks
in the emission weighted column shown in Fig.~\ref{fig:abs_col}. The
emission weighted column density is calculated as $\Sigma
\sigma_{\rm{tot}} I_{0} / \Sigma I_{0}$, where the summation is over
all sightlines to emitting regions and $\sigma_{\rm{tot}}$ is the
total column density (cm$^{-2}$) along each sightline. This weighting
is more informative than the column densities presented in
\cite{Lemaster:2007} which were only calculated along a single
sight-line into the system.

Resolution tests have determined that the minimum number of phase
steps required for convergence of the attenuated X-ray lightcurves is
dependant on the ratio $v_{\rm{orb}} / v_{\rm sl}$, with
of order 1000 phase steps required for an orbit with $e=0.9$ and
$v_{\rm{orb}} / v_{\rm sl} \approx 1$.

\subsection{Results}
\label{sec:results}

\subsubsection{The X-ray lightcurve}
\label{subsec:lightcurve}

In this section we compute X-ray lightcurves for the hypothetical
systems considered. For the O+O systems we use emissivity
and opacity data calculated assuming solar abundances
(Figs.~\ref{fig:emiss_spec} and ~\ref{fig:abs_spec} respectively). For
the WR wind we use data appropriate for WN8 abundances. 

\begin{figure}
\begin{center}
    \begin{tabular}{l}
      \resizebox{80mm}{!}{{\includegraphics{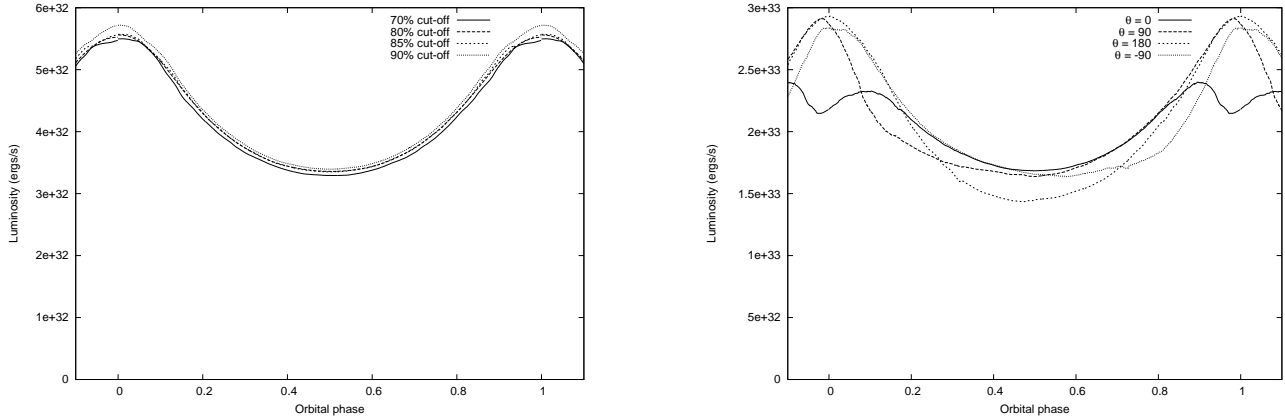}}} \\
    \end{tabular} 
\caption[]{Synthetic 0.1-10.0 keV lightcurves for different transition
positions between the shock cap and the point where the flow in the
WCR is assumed to behave ballistically for Model A (period of
$1\;$year, semi-major axis of $4.3\;$au). The transition point is
specified in terms of a percentage of the terminal speed of the slower
of the two winds, $v_{\rm sl}$. The assumed line-of-sight is
$i=90^{\circ}$ and $\theta = 0^{\circ}$. Interstellar absorption is
included.}
\label{fig:cutoff_lc}
\end{center}
\end{figure}

\begin{figure*}
\begin{center}
    \begin{tabular}{ll}
      \resizebox{80mm}{!}{{\includegraphics{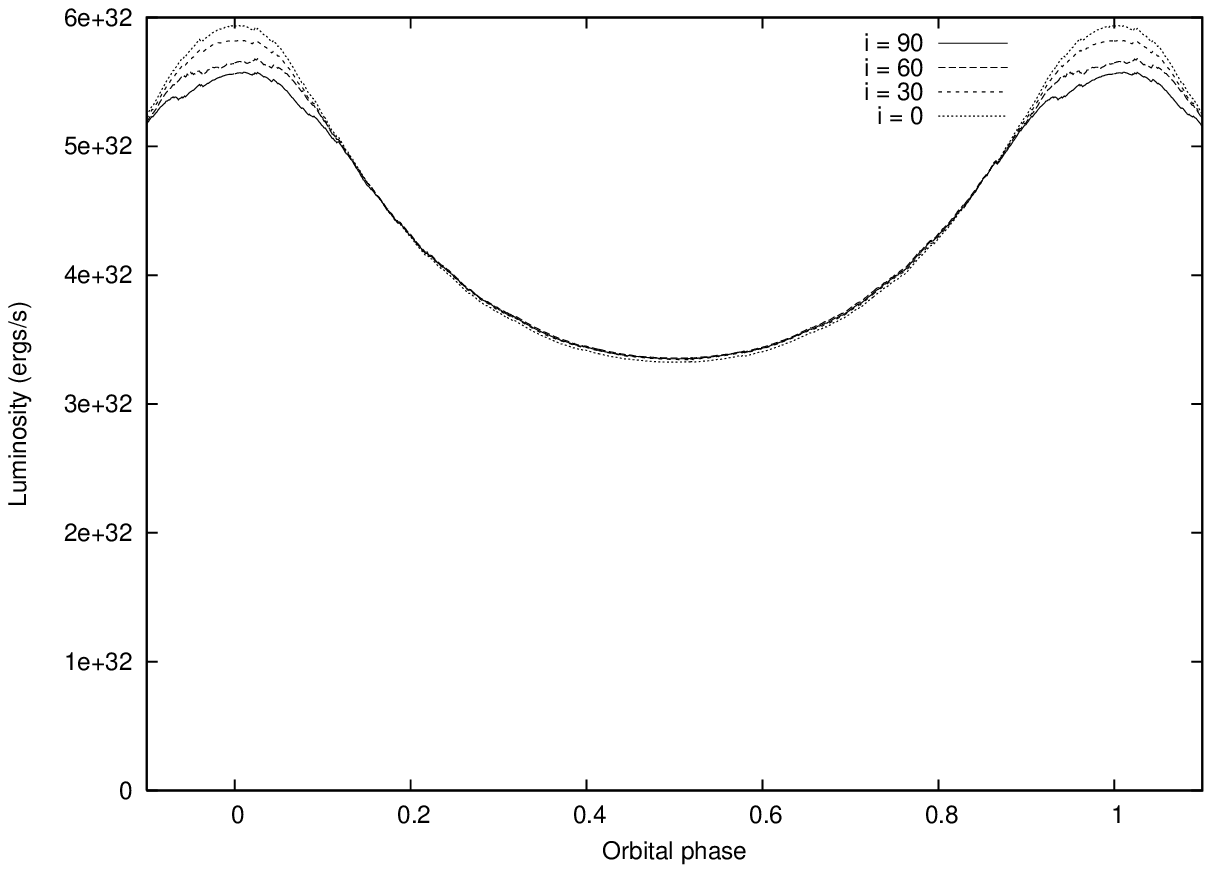}}} &
\resizebox{80mm}{!}{{\includegraphics{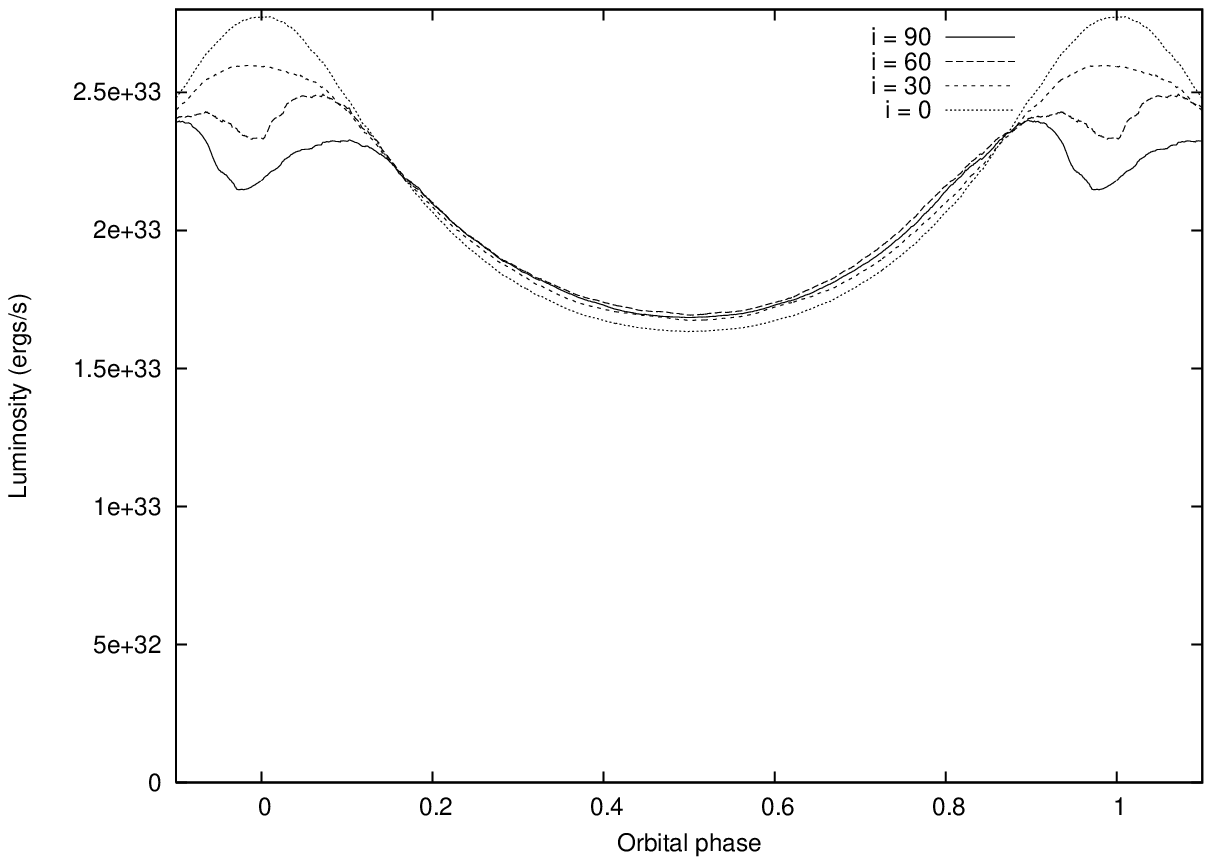}}} \\
    \end{tabular} 
\caption[]{Synthetic 0.1-10.0 keV emission lightcurves for various
inclination angles, $ i $, for Model A (left, $a = 4.3\;$au) and Model
B (right, $a = 0.81\;$au). Model parameters are indicated in
Table~\ref{tab:models}. Interstellar absorption is included.}
\label{fig:inc_lc}
\end{center}
\end{figure*}

Fig.~\ref{fig:cutoff_lc} shows the synthetic lightcurves produced for
models where the transition between the shock cap and the ballistic CD
is varied. There is a maximum divergence of $\sim 6\%$ between cases
where the transition occurs at a velocity cut-off of 70 \% and 90 \%
of the speed of the slower wind, which shows that the resulting
lightcurves are not very sensitive to this assumption.

Varying the orbital inclination angle changes the amount of
attenuation that the intrinsic emission suffers on its way to the
observer.  However, there is little circumstellar attenuation for
Model A (Fig.~\ref{fig:inc_lc}, left panel), and the synthetic
lightcurves are almost identical over the entire orbital period. This
is because the emitting volume is large (so occultation by the stars
is negligible), and because the stellar separation is wide, so that
the circumstellar density at the WCR is relatively low. Attenuation
effects become more prominent if the orbital period is reduced to
$1\;$month (Fig.~\ref{fig:inc_lc}, right panel), and distinct
differences in the lightcurves occur around periastron. The
$i=0^{\circ}$ lightcurve is smooth, and reflects the fact that the
increase in the intrinsic emission due to the changing orbital
separation ($L_{\rm x} \propto 1/d_{\rm sep}$) more than offsets the
peak in attenuation through the primary wind at periastron. Increasing
the inclination enhances the attenuation around periastron. The dip
seen in both the $i=60^{\circ}$ and $i=90^{\circ}$ curves is offset
from the time of periastron ($\phi = 1.0$) because of the skew to the
WCR caused by orbital motion. As already mentioned in
\S~\ref{subsec:occultation}, the stars fail to provide any significant
eclipse of the emitting region.

\begin{figure}
\begin{center}
    \begin{tabular}{l}
      \resizebox{80mm}{!}{{\includegraphics{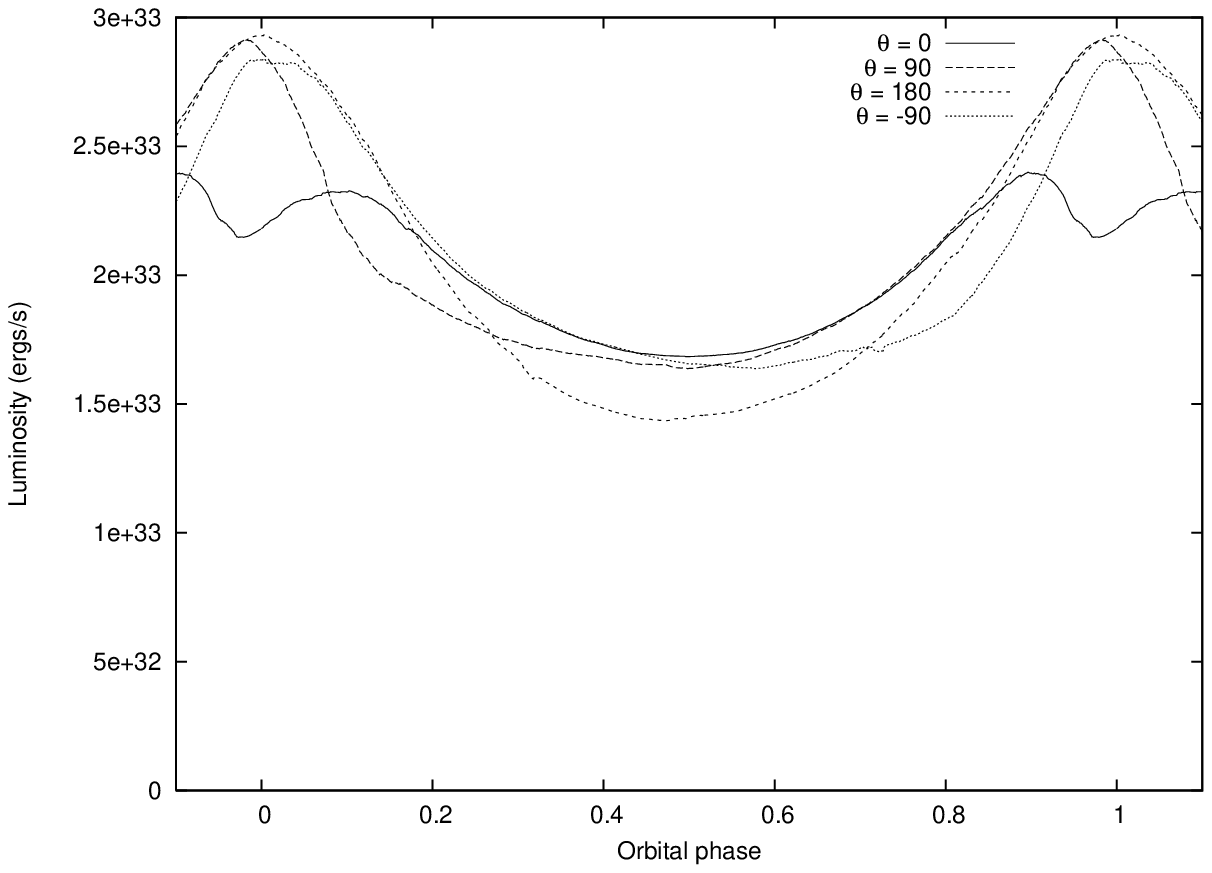}}} \\
    \end{tabular} 
\caption[]{Synthetic 0.1-10.0 keV emission lightcurves for Model B
(see Table~\ref{tab:models}) with $i = 90^{\circ}$ and various
line-of-sight angles, $\theta$. The observed emission is clearly
sensitive to the viewing angle. Interstellar absorption is included.}
\label{fig:los_lc_1month}
\end{center}
\end{figure}

Fig.~\ref{fig:los_lc_1month} examines the dependence of the observed
emission on the angle subtended between the line-of-sight and the
semi-major axis. The $\theta = 90^{\circ}$ and $-90^{\circ}$ curves
appear to be almost identical reflected copies around orbital phase
$\phi \simeq 0.5$, with the differences around periastron being due to
the aberration of the WCR. Absorption does not strongly affect the
observed emission (even if the orbital period is reduced to 1 month)
as the density contrast between the O-star winds is not very
large. The largest difference between the model results ($\sim 25\%$)
occurs at periastron. A comparison between current observational data
and such models may allow constraints to be placed on the orientation
of specific O+O-star systems.

\begin{figure}
\begin{center}
    \begin{tabular}{l}
      \resizebox{80mm}{!}{{\includegraphics{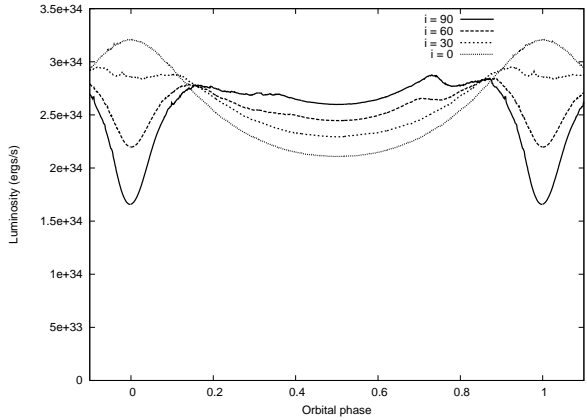}}} \\
    \end{tabular} 
\caption[]{Synthetic 0.1-10.0 keV emission lightcurves for Model C
(see Table~\ref{tab:models}) with various inclination angles, $ i
$. Interstellar absorption is included.}
\label{fig:inc_lc_wr}
\end{center}
\end{figure}

\begin{figure}
\begin{center}
    \begin{tabular}{l}
      \resizebox{80mm}{!}{{\includegraphics{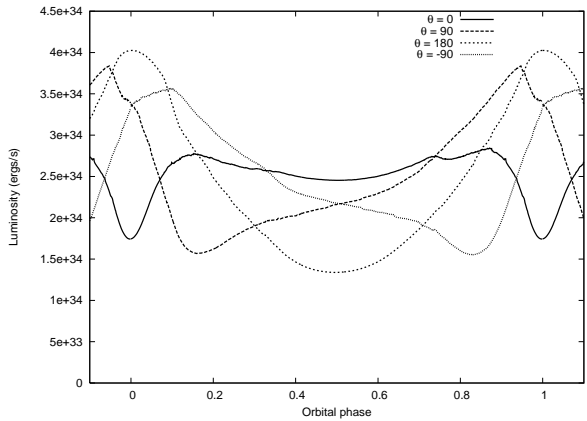}}} \\
    \end{tabular} 
\caption[]{Synthetic 0.1-10.0 keV emission lightcurve for Model C
(see Table~\ref{tab:models}) with inclination $i = 90^{\circ}$ and various
line-of-sight angles, $\theta$. Interstellar absorption is included.}
\label{fig:los_lc_wr}
\end{center}
\end{figure}

The higher primary mass-loss rate in the WR+O system leads to a
greater depenence of the observed emission on the line-of-sight
(Figs.~\ref{fig:inc_lc_wr} and ~\ref{fig:los_lc_wr}), as well as
higher X-ray luminosities. The minimums in the curves close to
periastron in Fig.~\ref{fig:inc_lc_wr}, especially in the case of the
$i=90^{\circ}$ curve, are the result of the the X-rays passing through
the dense WR wind. However, there is again little difference between
the $i=60^{\circ}$ and $i=90^{\circ}$ curves at apastron as the WCR is
viewed predominantly through the less dense O-star wind, though the
attenuation at lower inclinations is slightly higher as the apex of
the WCR is viewed through the denser wind from the WR star. Rotating
the line-of-sight within the orbital plane again causes significant
alterations to the observed emission (Fig.~\ref{fig:los_lc_wr}). The
$\theta=0^{\circ}$ curve sees the largest degree of attenuation around
periastron and the lowest around apastron, with the opposite being
true for the $\theta=180^{\circ}$ curve. As was also the case in
Fig.~\ref{fig:los_lc_1month}, the $\theta = 90^{\circ}$ and
$-90^{\circ}$ lightcurves appear to be almost perfect reflected copies
of each other. The dip seen in the $\theta = 90^{\circ}$ curve at
orbital phase $\phi \simeq 0.17$ marks opposition. Features such as
these could be particularly useful for constraining the orientation of
systems.

\begin{figure}
\begin{center}
    \begin{tabular}{l}
      \resizebox{80mm}{!}{{\includegraphics{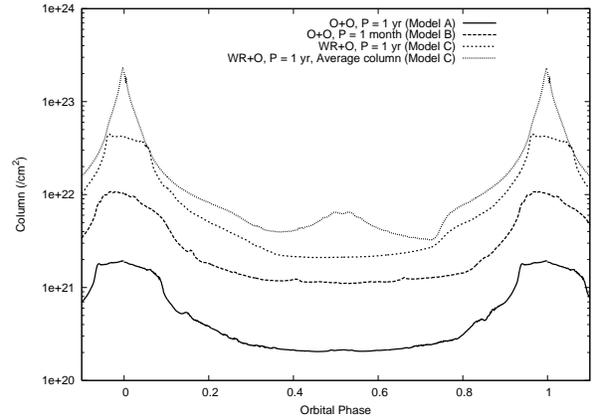}}} \\
    \end{tabular} 
\caption[]{Variation of the emission weighted column density as a
function of orbital phase for the O+O and WR+O-star systems
considered. The average column is also shown for the WR+O-star system
to demonstrate the difference between this and an emission weighted
column. Interstellar absorption is included.}
\label{fig:all_col}
\end{center}
\end{figure}

Fig.~\ref{fig:all_col} shows the variation with phase of the emission
weighted column density for the 3 hypothetical systems considered. The
column density is highest when viewed through the primary wind, and
lowest when viewing through the secondary wind. It is lowest for Model
A, and is approximately $5 \times$ higher when the period is reduced
to $1\;$month (Model B). This simply reflects the $\simeq 5 \times$
smaller separation and the $\simeq 30 \times$ higher densities. The
$\sim 10 \times$ changes in the column density between apastron and
periastron in Model A curve reflects the $1.86 \times$ change in
stellar separation and $5\times$ change in wind density (or stellar
mass-loss rate) as the line-of-sight switches from the secondary wind
into the primary wind.

The different slopes of the column density either side of periastron
are caused by the asymmetry of the WCR. The rise in column density at
$\phi \sim 0.8$ begins when the shock cap rotates and lines-of-sight
start to see the emission through the denser primary wind. The rise
occurs at an earlier phase for the WR+O system because of the lower
value of the wind momentum ratio and the narrowing of the opening
angle of the WCR. At $\phi = 0.964$ and 0.055 the slope in the
emission weighted column is reduced, and this feature marks the point
where the bow shock arms are tangential with the line-of-sight. When
the emission weighted column density is plotted alongside the average
column density the change in slope occurs at a point where the two
curves intersect. The average column density has a continual rise and
a peak at periastron. This tells us that the column density to the
entire emission region reaches a maximum at periastron, whereas
attenuation to the points with highest intrinsic emission remains
roughly constant for a short period. It is also interesting that both
the O+O and WR+O systems with $P = 1\;$year (Model A and C
respectively) have flatter profiles at maximum column density. This
indicates that the shape of the column density curve is sensitive to
the aberration and orbital induced curvature of the WCR, and thus to
the orbital period.

\subsubsection{X-ray spectra}
\label{subsec:spectra}

\begin{figure}
\begin{center}
    \begin{tabular}{l}
      \resizebox{80mm}{!}{{\includegraphics{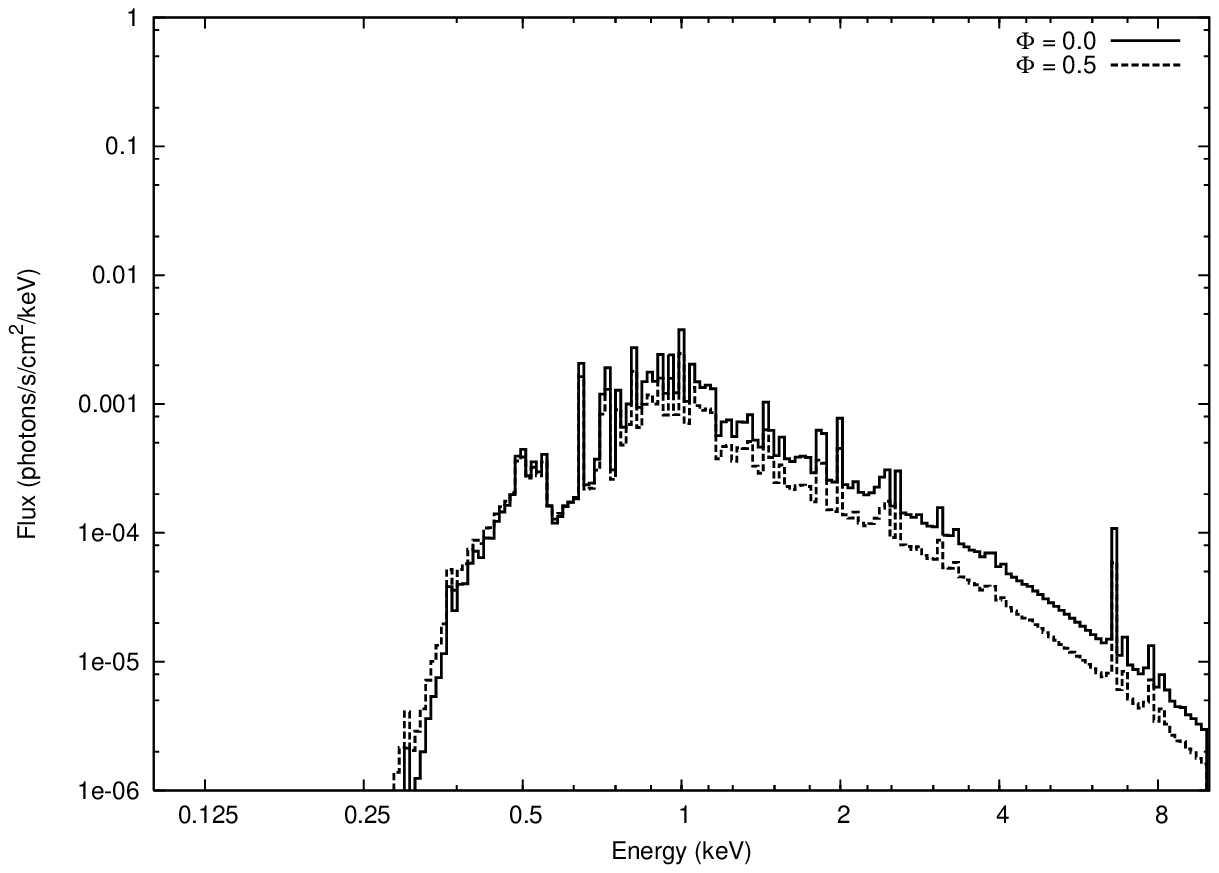}}} \\
      \resizebox{80mm}{!}{{\includegraphics{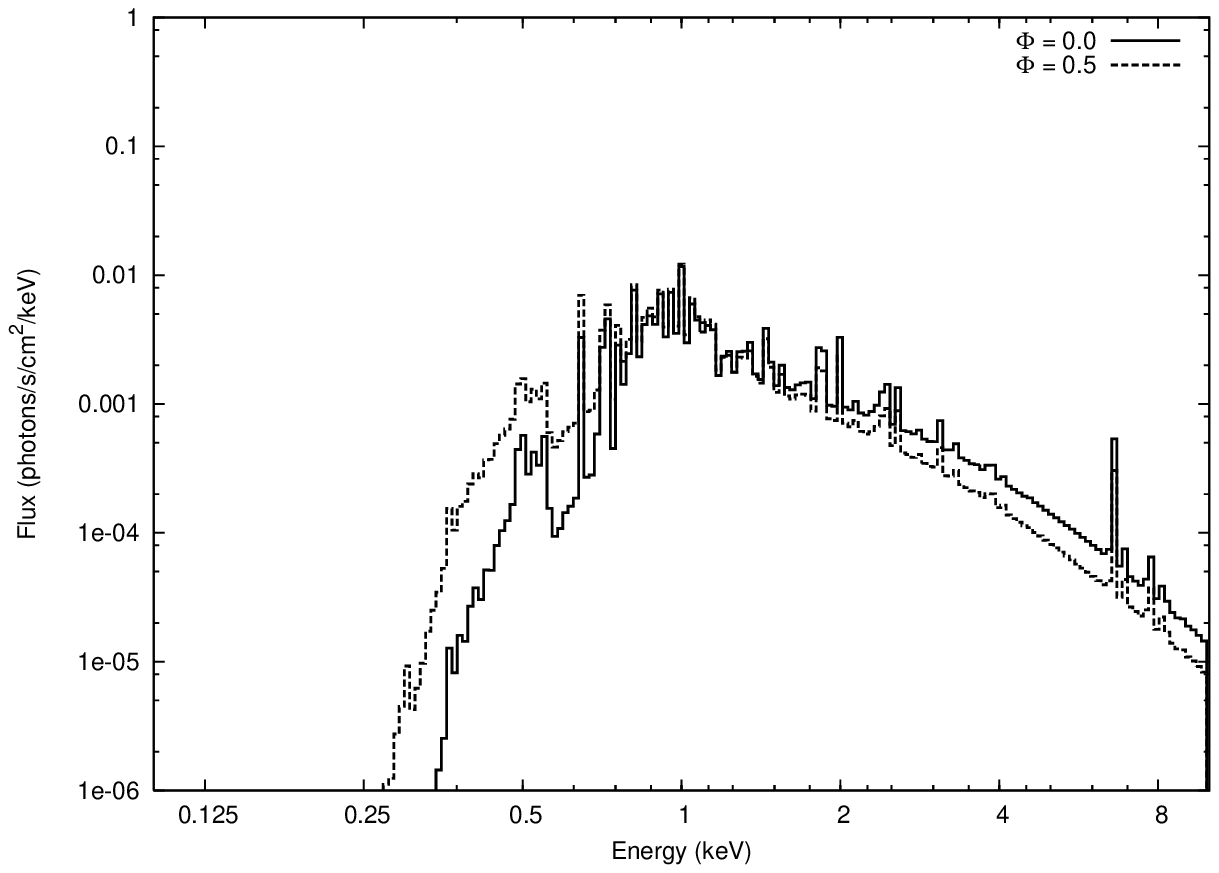}}} \\
      \resizebox{80mm}{!}{{\includegraphics{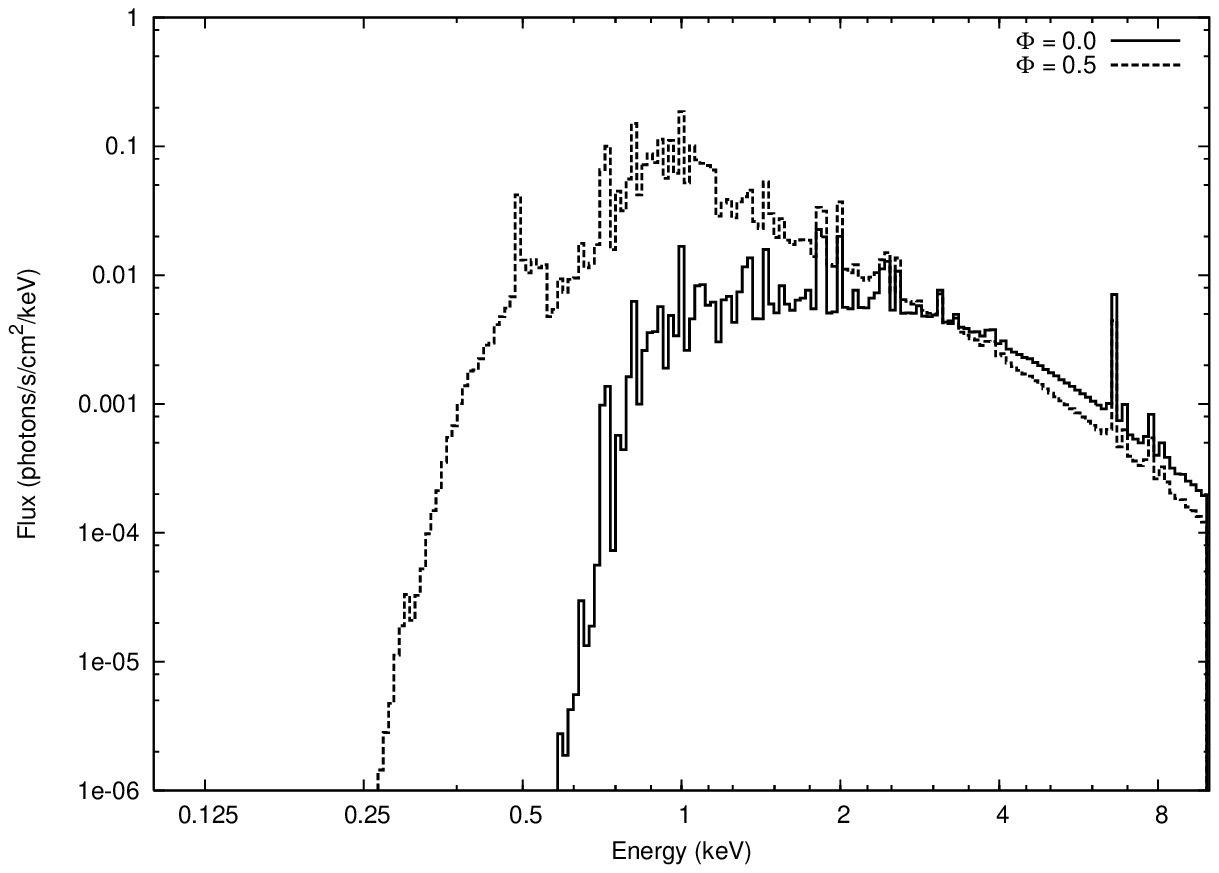}}} \\
    \end{tabular} 
\caption[]{Synthetic 0.1-10.0 keV spectra at periastron and apastron
  for Model A (top), B (middle), and C (bottom). The viewing angles
  are $i=90^{\circ}$ and $\theta = 0^{\circ}$. Interstellar absorption
  is included. The corresponding lightcurves for these spectra are
  shown in Figs.~\ref{fig:inc_lc}, ~\ref{fig:los_lc_1month} and
  ~\ref{fig:inc_lc_wr}.}
\label{fig:specs}
\end{center}
\end{figure}

Fig.~\ref{fig:specs} shows synthetic spectra at periastron and
apastron for the simulations performed. The slope of the spectra at
high energies is the same for the O+O systems since the preshock
velocities, and therefore postshock gas temperatures, do not
change. In all cases spectra at apastron show lower flux in the
2.0-10.0 keV energy band, although flux below 1 keV is higher. This is
because the intrinsic emission scales as $1 / d_{\rm{sep}}$, but the
is weaker when viewed through the companion's wind. The low energy
turnover in the periastron spectrum extends to higher energies for the
WR+O system due to the higher mass-loss rate and absorption of the WR
wind.

\section{Conclusions}
\label{sec:conclusions}

We have presented a 3D dynamical model of the colliding winds in
binary systems where both stars drive a significant wind. In circular
systems, the WCR adopts a spiral shaped structure similar to those
observed in massive binary star systems. In systems with eccentric
orbits, the shape of the WCR becomes increasingly deformed as the
eccentricity increases, with the winds increasingly being channeled
into a specific direction. A major advantage of the model is its low
computational cost and the fact that it can be easily adapted to model
a wide variety of observational data (from the radio to $\gamma$-ray)
and systems (from early type binaries, to $\gamma$-ray binaries with a
pulsar wind, to symbiotic novae).

As an example exercise, the X-ray emission from hypothetical O+O and
WR+O-star systems was modelled. The intrinsic emission was computed
from a 2D grid-based hydrodynamical model of the WCR, and then mapped
onto the surface separating the winds in the 3D dynamical model.
Absorption due to the unshocked stellar winds (and also cooled
postshock material) can be considered, although in the hypothetical
systems that were modelled the gas in the WCR remains largely
adiabatic as it flows out of the system so that only the former is
calculated. Ray-tracing through the 3D spiral structures then gives
the attenuated emission, and synthetic spectra and lightcurves are
produced.

The lightcurves and spectra show that observational characteristics of
the X-ray emission from early-type binaries can be reproduced. For
instance, the model with a 1 year orbit (Model A) is representative of
wide O+O binaries such as HD\thinspace15558 \citep{DeBecker:2006a},
and in this particular system could be useful in determining whether
there are two or three counterparts. The results from the 1 month
orbit simulation (Model B) are instead most applicable to X-ray
observations of close O+O binaries such as $\iota$ Orionis
\citep{Pittard:2000}, CygOB2\#8A \citep{DeBecker:2006b}, and HD 93403
\citep{Rauw:2000}, to name but a few. The model can also tackle
systems with different abundances for each wind such as WR+O-star
systems. Our WR+O star model with a $1\;$ year orbital period (Model
C) is applicable to systems like WR\thinspace25, WR\thinspace108,
WR\thinspace133, WR\thinspace138 \citep{vanderHucht:2001}, and among
the WN stars and WR\thinspace19, WR\thinspace125, WR\thinspace137,
WR\thinspace98a, WR\thinspace104, and WR\thinspace140
\citep{Pollock:2005,Pittard:2006} among the WC stars.

Mass-loss rate determinations can be made from comparison of the
predicted magnitude of the X-ray flux with observations
\citep{Stevens:1996,Pittard:2002}. In principle it is possible to use
the shape of the X-ray lightcurve to constrain the inclination and
orientation of the system. Our results reveal that for wind momentum
ratios of order 0.2, the lack of significant absorption means that
this will be very difficult if applied to O+O-star systems with
periods of order one year, but becomes possible for orbital periods of
order one month. The variation in absorption is much more significant
when the wind momentum ratio is lower and the density of the winds is
more disparate. This is the case for WR+O, LBV+O, and LBV+WR systems.

In future work we will apply the dynamical model to the X-ray and
forbidden line emission from $\eta\;$Car, the X-ray lightcurve of
WR\thinspace140, and emission line profiles of colliding wind
binaries.

\subsection*{Acknowledgements}
We would like to thank Perry Williams for the 2D code which was the
basis for the 3D model in this work. ERP thanks the
University of Leeds for funding. JMP gratefully acknowledges funding
from the Royal Society.


\label{lastpage}


\end{document}